\documentclass[aps, reprint, amsmath,amssymb, prl, superscriptaddress]{revtex4-2}
\usepackage{svg}
\usepackage{color}
\usepackage{graphicx}
\usepackage{times}
\usepackage{dcolumn}
\usepackage{bm}
\usepackage{physics}
\usepackage{xr-hyper}
\usepackage{hyperref}
\usepackage{url}
\usepackage{amsmath}
\usepackage{tikz-cd}
\usepackage{notes2bib}
\usepackage{multirow}
\usepackage{mathrsfs}
\usepackage{tabularx}
\usepackage{ dsfont }
\usepackage{svg}
\usepackage{verbatim}
\usepackage{amssymb}
\usepackage{amsfonts}              
\usepackage{amsthm}                
\usepackage{mathtools}
\usepackage{qcircuit}
\usepackage[normalem]{ulem}
\usepackage{xcolor}
\usepackage{csquotes}
\usepackage[many]{tcolorbox}
\usetikzlibrary{arrows.meta, calc}

\usepackage[clock]{ifsym}
\usepackage{fontawesome}
\usepackage{manfnt}

\usepackage{enumitem}

\DeclarePairedDelimiterX{\inp}[2]{\langle}{\rangle}{#1, #2}

\newcommand{\id}{\text{id}}

\newcommand{\swap}{{\tt SWAP}}

\makeatletter
\newcommand*{\addFileDependency}[1]{
  \typeout{(#1)}
  \@addtofilelist{#1}
  \IfFileExists{#1}{}{\typeout{No file #1.}}
}
\makeatother

\makeatletter
\newcommand{\externaldocumentNoBib}[1]{%
  \begingroup
    \let\bibcite\@gobbletwo 
    \externaldocument{#1}%
  \endgroup
}
\makeatother

\DeclarePairedDelimiterX{\barpair}[2]{(}{)}{%
  #1\;\delimsize\|\;#2%
}

\usepackage{stackengine,scalerel}

\theoremstyle{definition}
\newtheorem{theorem}{Theorem}
\newtheorem{prop}{Proposition}

\newtheorem{example}{Example}
\newtheorem{definition}{Definition}

\newtheorem{exercise}{Exercise}

\usepackage[normalem]{ulem}
\bibnotesetup{note-name    = ,use-sort-key = false}

\newcommand{\mds}[1]{\mathds{#1}}

\newcommand{\jami}{Jamio{\l}kowski }

\newcommand{\cA}{\mathcal{A}}

\newcommand{\cD}{\mathcal{D}}
\newcommand{\cE}{\mathcal{E}}
\newcommand{\cF}{\mathcal{F}}

\newcommand{\cH}{\mathcal{H}}
\newcommand{\cI}{\mathcal{I}}

\newcommand{\cM}{\mathcal{M}}

\newcommand{\cS}{\mathcal{S}}

\newcommand{\cU}{\mathcal{U}}

\newcommand{\cW}{\mathcal{W}}
\newcommand{\cX}{\mathcal{X}}
\newcommand{\cY}{\mathcal{Y}}
\newcommand{\cZ}{\mathcal{Z}}

\newcommand{\fI}{\mathfrak{I}}

\begin{document}
\title{Probing Quantum States Over Spacetime Through Interferometry}

\author{Seok Hyung Lie}
\email{seokhyung@unist.ac.kr}
\affiliation{
 Department of Physics, Ulsan National Institute of Science and Technology (UNIST), Ulsan 44919, Republic of Korea
}%

\author{Hyukjoon Kwon}
\affiliation{
 School of Computational Sciences, Korea Institute for Advanced Study, Seoul 02455, Republic of Korea
}

\begin{abstract}
Establishing a notion of the quantum state that applies consistently across space and time could be a crucial step toward formulating a relativistic quantum theory. We give an operational meaning to multipartite quantum states over arbitrary regions in spacetime through a \textit{causally agnostic measurement}, a measurement scheme that can be consistently implemented independently of the causal relation between the regions. We prove that such measurements can always be implemented with interferometry, also known as the scattering circuit technique, wherein the conventional density operator, the recently developed quantum state over time (QSOT), and the process matrix formalisms smoothly merge. This framework allows for a systematic study of mixed states in the temporal setting, which turn out to be crucial for modeling quantum non-Markovianity. Based on this, we demonstrate that two different ensembles of quantum dynamics can be represented by the same QSOT, indicating that they cannot be distinguished through interferometry. Moreover, our formalism reveals a new type of spatiotemporal correlation between two quantum dynamics that originates from synchronized propagation in time under time-reversal symmetry. We show that quantum systems with such correlation can be utilized as a reference frame to distinguish certain dynamics indistinguishable under time-reversal symmetry.
\end{abstract}

\maketitle
The observer dependence of spacetime in relativity calls for a unified description of quantum systems in spacetime. This led to several higher-order process theories such as process matrices~\cite{oreshkov2012quantum,araujo2014quantumcontrol,procopio2015suporder,araujo2015witness,oreshkov2016causal}, process tensors~\cite{pollock2018completeframework,pollock2018operationalmarkov,jorgensen2019tensor,white2020demonstration,milz2019cpdivisibility} and quantum combs~\cite{chiribella2013qswitch,chiribella2009framework,chiribella2010purification,chiribella2008architecture,chiribella2008supermaps}, which encode how spatiotemporal quantum processes respond to arbitrary (often counterfactual) interventions of an experimenter. However, these frameworks still treat temporal correlations as fundamentally different from spatial ones: there is no prescription for representing time-separated correlations as a bona fide quantum state. Consequently, a truly symmetric, state-based framework over spacetime remains elusive.

To address this problem, a \emph{quantum state over time} (QSOT) formalism has been developed~\cite{accardi2020quantum, leifer2013towards, fullwood2022quantum, fullwood2023quantum, parzygnat2023time, aharonov2014each, APT10, APTV09}, representing temporal correlations with operators analogous to density matrices. Building on the pioneering work~\cite{leifer2013towards}, successive proposals~\cite{horsman2017can, fullwood2022quantum, milz2021quantum} culminated in the uniquely characterized Fullwood-Parzygnat product~\cite{fullwood2022quantum, lie2023uniqueness, parzygnat2023virtual, lie2024unique}, which reduces to the pseudo-density operator for multi-qubit systems~\cite{fitzsimons2015quantum, marletto2021temporal}. Leveraging its strong connection to quasiprobability distributions~\cite{kirkwood1933quantum, Lostaglio2023kirkwooddirac, abbott2019anomalous, margenau1961correlation} and weak measurements~\cite{aharonov1988result, lundeen2012procedure}, the QSOT formalism has enabled advances in metrology~\cite{arvidsson2020quantum}, Bayesian thermodynamics~\cite{parzygnat2023time, buscemi2021fluctuation}, entanglement in time~\cite{milekhin2025observable, Ghodarti2025String} and quantum transport~\cite{hoogsteder2025approach}. Despite this progress, QSOTs still lack a universally applicable, operational measurement scheme consistent with conventional quantum states: Existing protocols based on weak measurements~\cite{Mitchison2007sequential, lundeen2012procedure, Halpern2018quasiprobability, abbott2019anomalous, Lostaglio2023kirkwooddirac}, interferometric techniques~\cite{Mazzola2013measuring, Pedernales2014efficient, Halpern2017Jarzynski} and quantum snapshotting~\cite{Wang2024snapshotting} depend on a presumed causal order and fail to establish a one-to-one correspondence between outcomes and the underlying QSOT.

\begin{figure}
    \includegraphics[width=\linewidth]{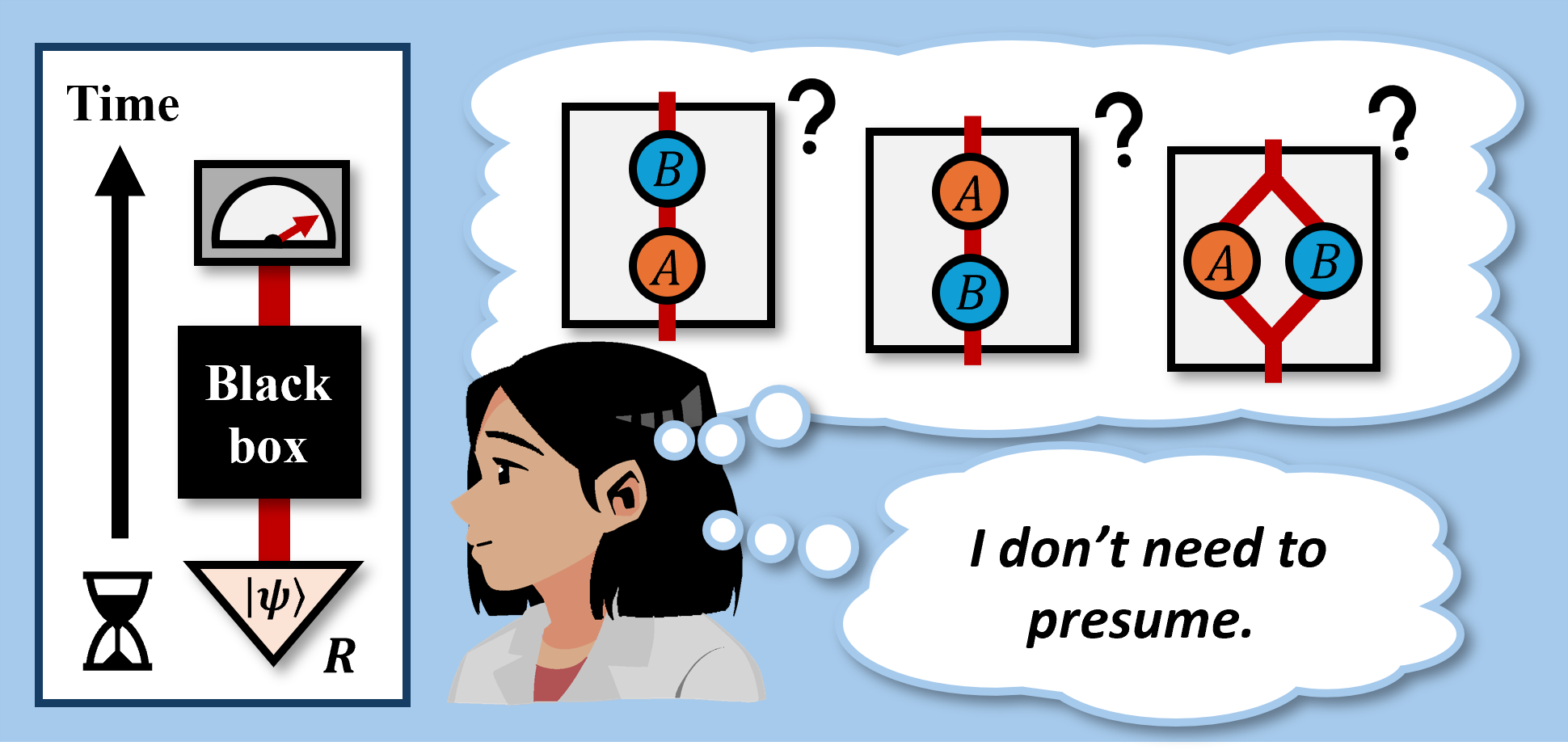}
    \caption{\textit{Causally agnostic measurement}. A probe $R$ interacts with an unknown process (black box) involving systems $A$ and $B$ localized in spacetime, then is measured. The interactions between the black box and $R$ must be invariant under all possible causal relations of $A$ and $B$ (top-right).
    }
   \label{fig:agnostic}
\end{figure}

In this work, we introduce \textit{causally agnostic} quantum measurements that do not require measurement devices to have any knowledge of the underlying causal structure, applicable to arbitrary regions in spacetime (see Fig.~\ref{fig:agnostic}).
We prove that the necessary and sufficient condition for such causally agnostic measurements is implementability via interferometry. We then show that the interference term in the outcome probabilities of interferometry completely characterizes the quantum state over spacetime with a clear operational meaning that unifies the density operator and the QSOT formalisms.

Remarkably, our quantum state over spacetime formalism reveals a genuine multipartite spatiotemporal correlation between parallel quantum dynamics that emerges under time‐reversal symmetry, which we term \emph{synchronization}. We construct examples of QSOTs that contain less information under time-reversal symmetry 
and show that such an information deficit can be removed by appending a reference qubit that co‑propagates with the system in time, i.e., synchronized with the dynamics of interest, while carrying no other dynamical data.  This demonstrates how a temporal quantum reference frame can distinguish future from past even when the underlying dynamics are time‐reversal symmetric. Being readily implementable with interferometers, our formalism offers immediate application for exploring spatiotemporal correlations of various quantum systems.

\textit{QSOT products---}
Let a pair $(\rho,\cE)$ of an initial state $\rho_A$ and a channel $\cE:A \to B$ be referred to as a \textit{dynamics}  (from $A$ to $B$). In previous works ~\cite{accardi2020quantum, leifer2013towards, fullwood2022quantum, fullwood2023quantum, parzygnat2023time, aharonov2014each, APT10, APTV09}, a QSOT $\cE\star \rho$ has been constructively defined for a given dynamics $(\rho,\cE)$ as an operator on ${\cal H}_A \otimes {\cal H}_B$, where ${\cal H}_A$ and ${\cal H}_B$ are the Hilbert spaces of the systems $A$ and $B$, respectively. Similarly to the reduced density matrices of a multipartite quantum system, a QSOT product is required to satisfy the marginality condition,
\begin{equation}
    \label{eq:marginal_cond}
    \Tr_A[\cE\star\rho ] = \cE(\rho) \;\text{and}\; \Tr_B[\cE\star\rho] = \rho.
\end{equation}
We define three popular bipartite QSOT products of main interest in this work. Let $J[\cE]$ be the \jami operator of $\cE$ defined as $J[\cE]:= \sum_{i,j}\dyad{i}{j}_A \otimes \cE(\dyad{j}{i})_{B}$. The \textit{left product} $\star_L$ is defined as $\cE \star_L \rho = (\rho_A \otimes \mds{1}_B) J[\cE]$. Similarly, the \textit{right product} $\star_R$ is defined as \(\cE \star_R \rho = J[\cE] (\rho_A \otimes \mds{1}_B)\). The \textit{symmetric product}, or the \textit{Fullwood-Parzygnat (FP) product} $\star_{FP}$ is defined as the average of the products above:
\begin{equation}
    \cE \star_{FP} \rho = \frac{1}{2}(\cE \star_L \rho + \cE \star_R \rho).
\end{equation}

\textit{Causally agnostic measurements}--- 
Let us review the scheme of general indirect measurement of $n$ subsystems $A_1,A_2,\dots,A_n$~\cite{wilde2013quantum}: a probe $R$ is prepared in their joint past and interacts with $A_i$ for each $i=1,\dots,n$ in some order (or even simultaneously), and is finally measured in their joint future. In a narrow sense, the set of admissible interactions between systems and the probe,  characterized by unitary operators $U_{A_i R}$ defines a measurement scheme.

However, in many cases, one can only know what kind of interaction could take place between each system and the probe, but not when and where. For example, in the Glauber multiple scattering theory ~\cite{glauber1970scattering,boal1990glauber,franco1966deuteron,franco1972glauber,bertulani1988electromagnetic}, collisions between incident and target particles contribute to the probability amplitude independently of their time ordering. Therefore, it is desirable to characterize the information that is extractable independently of the causal structure of the interaction. We call a measurement scheme \emph{causally agnostic} if every admissible interaction between the probe and each subsystem is invariant under any ordering, so that the composite system $A_1\cdots A_n$ can be treated as a black box with no assumed causal relation between subsystems $A_i$. 

\begin{theorem}\label{thm:CAM}
A quantum measurement is causally agnostic if and only if it can be implemented by a (multi-arm) \emph{interferometer}.
\end{theorem}

\begin{figure}
    \centering
    \includegraphics[width=\linewidth]{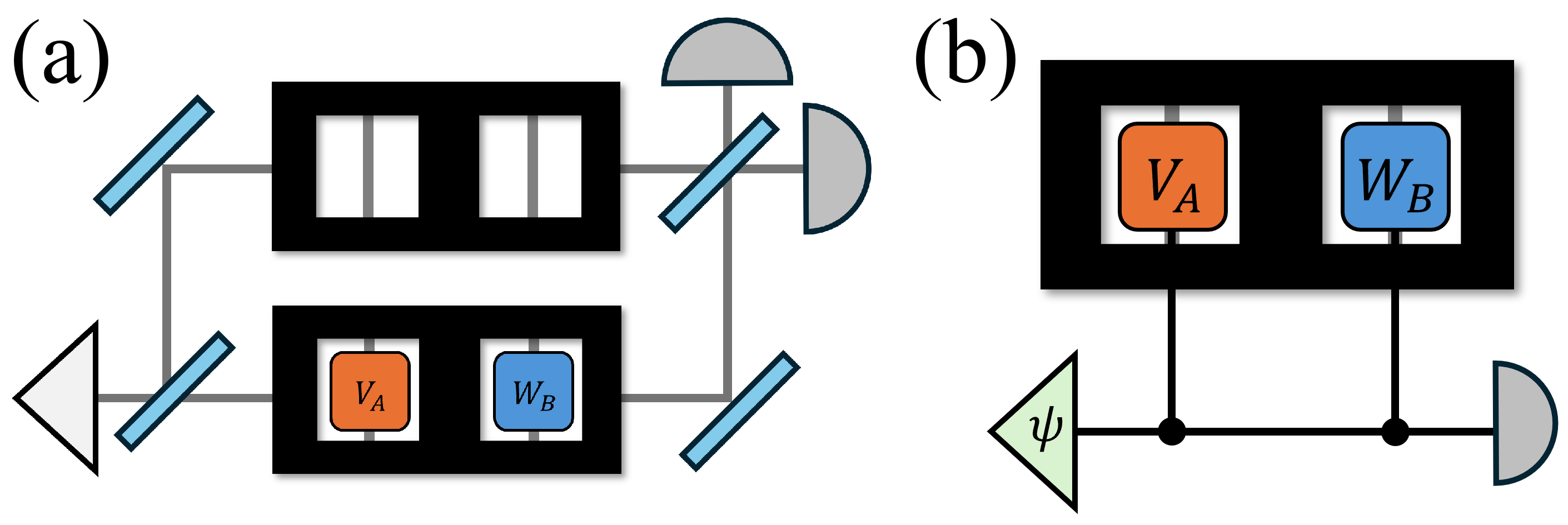}
    \caption{\textit{Interferometry of causal black box}. (a) One can probe a causal black box by measuring the interference between intervened and non-intervened ones. (b) The same interferometry can be implemented with a scattering circuit with controlled unitaries with an unspecified ordering.}
    \label{fig:interf}
\end{figure}

The key proof idea is that, in such a measurement scheme, interactions of $A_i$ with the probe system $R$ must commute on $R$ so that their ordering is irrelevant. See Sec. \ref{SM:CharInt} of \cite{SM} for proof. Theorem \ref{thm:CAM} not only characterizes interferometry---a versatile and readily deployable method with a wide variety of probe systems, such as spatial (Mach–Zehnder \cite{Mach1892,Zehnder1891}), internal (Ramsey \cite{Ramsey1950}) degrees of freedom---as the only kind of causally agnostic measurement scheme (see Fig.~\ref{fig:interf}), but also generalizes interferometry to unconventional settings where one does have to presuppose the causal structure of the measured system. An immediate consequence of Theorem \ref{thm:CAM} is that any measurement that cannot be implemented by interferometry requires a spatiotemporal \emph{reference frame} for its implementation; for example, a clock that tells us whether Alice interacts with the probe before or after Bob does, or a map that specifies the subsystem of the probe interacting with Alice and Bob. Since every multi-arm interferometer can be viewed as a statistical mixture of two-arm interferometers~\cite{sorkin1994quantum}, ``interferometry” in this work refers to two-arm interferometry unless otherwise stated.

\textit{Interferometric characterization of quantum states---}
A quantum state $\rho$ encodes probabilities of measurement outcomes $i$ for any measurement given as a POVM $\{M_i\}$ through the Born rule $\Pr(i) \;=\; \Tr\!\bigl[\rho\,M_i\bigr]$. Requiring $\Pr(i)$ to be non-negative for all positive operators $M_i$ then yields the conventional positivity condition on density operators. However, such a characterization has resisted a smooth generalization to the temporal setting due to apparent peculiarities of the QSOTs, such as negative or complex eigenvalues~\cite{fitzsimons2015quantum} and the effect of 
measurement back action, such as state disturbance, which is inevitable in quantum mechanics \cite{Heisenberg1927, Fuchs1996, LeggettGarg1985, Banaszek2001, Buscemi2014, WisemanMilburn2010, BuschLahtiMittelstaedt1996} because it persists in time so that it alters the original dynamics one wanted to probe. 

To overcome these issues, we propose an alternative characterization of quantum states through interferometry.
Assume that a system $S$ is in state $\rho$, whose purification is $\ket{\phi_\rho}_{SE}$ with an external system $E$. A probe state initially prepared in $\ket{\psi}_R=\alpha_0 \ket{0}_R+\alpha_1 \ket{1}_R$ enters an interferometer and creates the path-superposed state.
On one arm, a unitary operator $V$ is applied to $S$, so that the entire state evolves into $\alpha_0\ket{\phi_\rho}_{SE}\ket0_R + \alpha_1 V_S\ket{\phi_\rho}_{SE}\ket1_R$. Finally, measuring the probe system $R$ in the basis $\{\ket{b_+},\ket{b_-}\}$ yields outcome probabilities
\begin{equation} \label{eqn:intertheta}
    \Pr(\pm) = \cS_\pm + 2\Re [\cA_\pm \, \cI].
\end{equation}
Here, $\cS_\pm = |\alpha_0 \braket{b_\pm}{0}|^2 + |\alpha_1 \braket{b_\pm}{1}|^2$ and $\cA_\pm = \alpha_0^*\alpha_1 \braket{0}{b_\pm} \braket{b_\pm}{1}$ are solely determined by the initial probe state and measurement basis \footnote{Note that the maximum visibility $\cS_\pm = 2\cA_\pm = 1/2$ can be achieved by appropriate choices of the probe state and measurement basis.}, while all the information about $\rho$ is contained in the \textit{interference term},
$$
\cI=\Tr[V\rho].
$$
Consequently, to have the interference term for all possible unitary operators $V$ is equivalent to a complete description of $\rho$, providing another complete characterization of the quantum state. 

Such a characterization can be readily applied to the case where $S$ is composed of spatially separated subsystems, say, $A$ and $B$ in state $\rho_{AB}$. In this case, if the interferometric interventions independently applied to $A$ and $B$ are given as unitary operators $V_A$ and $W_B$, then the interference term $\cI$ is calculated as 
\begin{equation} \label{eqn:spatinterf}
    \cI = \Tr[(V_{A}\otimes W_{B})\rho_{AB}].
\end{equation}

Now, what if $A$ and $B$ are in a causal relation, say, a dynamics $(\rho, \cE)$ from $A$ to $B$? Then the non-intervened dynamics $(\rho,\cE)$ unfolds on one arm, and the intervention unitaries $V_A$ and $W_B$ are applied before and after $\cE$ on the other arm of the interferometer. The interference term is given as $\cI=\Tr[W\cE(V\rho)]$ in this case (See End Matter for the detailed derivation.), which is also known as a correlation function in quantum field theory. Remarkably, the interference term can be re-expressed in terms of QSOT as follows: 
\begin{equation} \label{eqn:tempinterf}
    \cI = \Tr[(V_A\otimes W_B)(\cE\star_L \rho)].
\end{equation}
Comparing \eqref{eqn:spatinterf} and \eqref{eqn:tempinterf}, the QSOT $\cE\star_L \rho$ has a completely consistent interpretation with that of $\rho_{AB}$ in \eqref{eqn:spatinterf}, bolstering the status of QSOTs as quantum states in the temporal regime. This interpretation of QSOT resolves a common criticism about its nonpositive eigenvalues; as long as the overall measurement probabilities \eqref{eqn:intertheta} are nonnegative, the interference term \eqref{eqn:tempinterf} could take any complex value, as $\Tr[V\rho]$ does even for conventional density operators $\rho$. Given that causally agnostic measurements consistently provide a complete characterization of both $\rho_{AB}$ and $\cE \star_L \rho$, we define the quantum state over spacetime for general multipartite settings as follows (See Fig.~\ref{fig:interf}):
\begin{definition} \label{def:SOST}
    We say that $n$ systems $A_1,A_2,\dots,A_n$ are in the \textit{quantum state over spacetime} (QSOST) $\rho_{A_1\dots A_n}$ if the interference term $\cI$ of interferometry is $\Tr[(V_{A_1}\otimes \cdots \otimes V_{A_n})\rho_{A_1\dots A_n}]$ when the unitary intervention at $A_i$ is given as $V_{A_i}$ for $i=1,\dots,n$.
\end{definition}
\noindent
Since this definition is consistent with the previous constructive definition of QSOT and subsumes it, hereafter, we will refer to a QSOST compatible with time-like separated systems $A_i$ as a QSOT to distinguish it from conventional density operators, or a quantum state over space~\footnote{Note that, however, QSOTs and density operators are not mutually exclusive as one QSOST can be compatible with both time-like and space-like separations.}.

Since there exist unitary bases of the operator space, one can explicitly reconstruct $\rho_{A_1\dots A_n}$ from the observed interference term $\cI$.
This definition mitigates key limitations in existing frameworks for temporal quantum correlations. Pseudo-density operators (PDOs), while successful for multi-qubit systems, enjoy a transparent statistical interpretation only for ``light‑touch" observables with eigenvalues $\pm\lambda$~\cite{Liu2025QuantumCausal, Parzygnat2024TimeSym, Song2025CausalExplain}, with arbitrary‑dimensional extensions still emerging~\cite{Fullwood2024Operator}.  By contrast, QSOTs apply to arbitrary finite dimensions but have so far relied on indirect probing through quasiprobability such as quantum snapshotting~\cite{Wang2024snapshotting}; Definition~\ref{def:SOST}, realized through the interferometric protocol characterized by Theorem~\ref{thm:CAM}, provides an alternative valid for arbitrary unitary operators $V_i$ including the light-touch observables of all finite-dimensional systems.

We remark that the interferometric protocol for probing QSOST can be implemented with a `scattering circuit' that has been already implemented on various quantum platforms including NMR~\cite{Du2006ScatteringCircuitNMR, Batalhao2014WorkNMR}, trapped ions~\cite{An2015JarzynskiIon},~cold atoms \cite{Cerisola2017WorkMeterAtomChip}, superconducting platforms~\cite{OftelieCampisi2025OpenWorkQST}, neutral atoms~\cite{Bluvstein2022NeutralAtomProcessor} and photonic systems~\cite{Lee2013NonlinearFunctionalsPhotonics, Starek2018FredkinNPJQI}. Especially, probing temporal quantum correlation using NMR was also done in Ref. ~\cite{liu2024NMR} although not in a fully causally agnostic fashion. (See Sec.~\ref{SM:Feasible} of ~\cite{SM} for the relevance of scattering circuit.)

One may wonder how QSOST connects to process matrices~\cite{oreshkov2012quantum}. 
We also show that QSOST is the \textit{first-order approximation} of the process matrix with respect to the expansion of a weak measurement. (See Sec.\ref{SM:QSOTPM} of \cite{SM} for details.)
This observation explains their significant difference in mathematical complexity: an $n$-step QSOT on a $d$‑level system contains $d^{2n}$ parameters, whereas, for example, the corresponding process tensor requires $d^{4n}$. It is because the former models the unperturbed configuration of quantum systems in spacetime ``as is,'' whereas the latter encodes statistical behavior of the systems under arbitrary measurements that could lead to substantial modification of the original configuration due to measurement back action.

\textit{Mixed states over spacetime---} 
Consider a probabilistic mixture of dynamics $(\rho_i,\cE_i)$, each chosen with probability $p_i$.  Eq.~\eqref{eqn:tempinterf} suggests that the interference term becomes $\cI = \sum_i p_i \Tr \bigl[(V_A\otimes W_B)(\cE_i \star_L \rho_i)\bigr]$ which simplifies into
\[
 \Tr\!\qty[(V_A\otimes W_B)\Bigl(\sum_i p_i \cE_i \star_L \rho_i\Bigr)].
\]
According to Definition \ref{def:SOST}, such systems $A$ and $B$ are in the QSOT  $\rho_{AB}= \sum_i p_i \cE_i \star_L \rho_i$ in a \emph{mixed} form~\footnote{One may want to call a QSO(S)T that cannot be written as such a nontrivial convex combination of other QSO(S)Ts may be called \emph{pure}, but there are subtleties. See End Matter for a detailed discussion.}.

We define a QSOST to be \emph{factorizable} if it can be expressed as $\mathcal{E}\star\rho$ for some dynamics $(\rho,\mathcal{E})$ so that it is also a QSOT. With this definition, such mixed states are a remarkable example of non-factorizable QSOSTs.
While earlier works focused almost exclusively on factorizable QSOTs, the non-factorizability of QSOTs is essential as it provides a much more economical definition of non-Markovianity of quantum processes~\cite{lie2024unique}. (See Sec. \ref{SM:Markov} of~\cite{SM} for a more detailed discussion on factorizability.)

We remark that several distinct ensembles of dynamics can yield the same mixed state (see Example~\ref{ex:ensemble}) and thus become indistinguishable through interferometry. This is analogous to the fact that a density matrix admits many different pure‑state decompositions, and preferring one ensemble over another risks the preferred ensemble fallacy~\cite{nemoto2002partition}.

\textit{Time-reversal symmetry---} Usually, time-reversal asymmetry is taken for granted because almost always an experimenter is equipped with temporal reference frames, e.g., a clock on the wall, raindrops falling down outside, or her biological clock. However, like many other fundamental theories of nature, quantum mechanics, in principle, is symmetric under time-reversal because the closed system time evolutions are reversible, and one cannot prefer one temporal direction over the other without a temporal reference frame.

Thus, let us consider an interferometric protocol for a dynamics $(\rho,\cE)$ from $A$ to $B$ with time-reversal symmetry. Even when $\cE$ is irreversible, one always regards it a part of a reversible closed system evolution $\cU$ from $AE_A$ to $BE_B$ by considering its dilation  $\cE(\rho)=\Tr_{E_B}[\cU(\rho_A\otimes \tau_{E_A})]$ with auxiliary systems $E_A$ and $E_B$ at two different times of $A$ and $B$. Under the symmetry, the dilated dynamics $(\rho_A\otimes\tau_{E_A},\cU)$ can be equivalently described as $(\cU(\rho_A\otimes\tau_{E_A}),\cU^\dag)$ with the direction of time inverted. As they cannot be distinguished under the symmetry, the interference term given as \eqref{eqn:tempinterf} should be given as the equal mixtures of those associated with the respective dynamics, which is calculated to be $\cI = \Tr[W\cE(V\rho)+W\cE(\rho V)]/2$, (See Sec. \ref{SM:DervIntTRev} of \cite{SM}.) independent of the choice of unitary dilation. By observing that $\Tr[W\cE(\rho V)]=\Tr[(V_A\otimes W_B)\,( \cE\star_R \rho)]$, the interference term now becomes
\begin{equation} \label{eqn:FPprob}
   \cI=\Tr[(V_A\otimes W_B)\,(\cE \star_{FP} \rho)].
\end{equation}
In other words, if we assume time-reversal symmetry, which is common in microscopic processes, the Fullwood–Parzygnat product arises naturally. This product coincides with the PDO formalism in special cases and was uniquely characterized in Ref.~\cite{lie2023uniqueness} from physically motivated axioms, including time-reversal symmetry.

\begin{figure}
    \includegraphics[width=\linewidth]{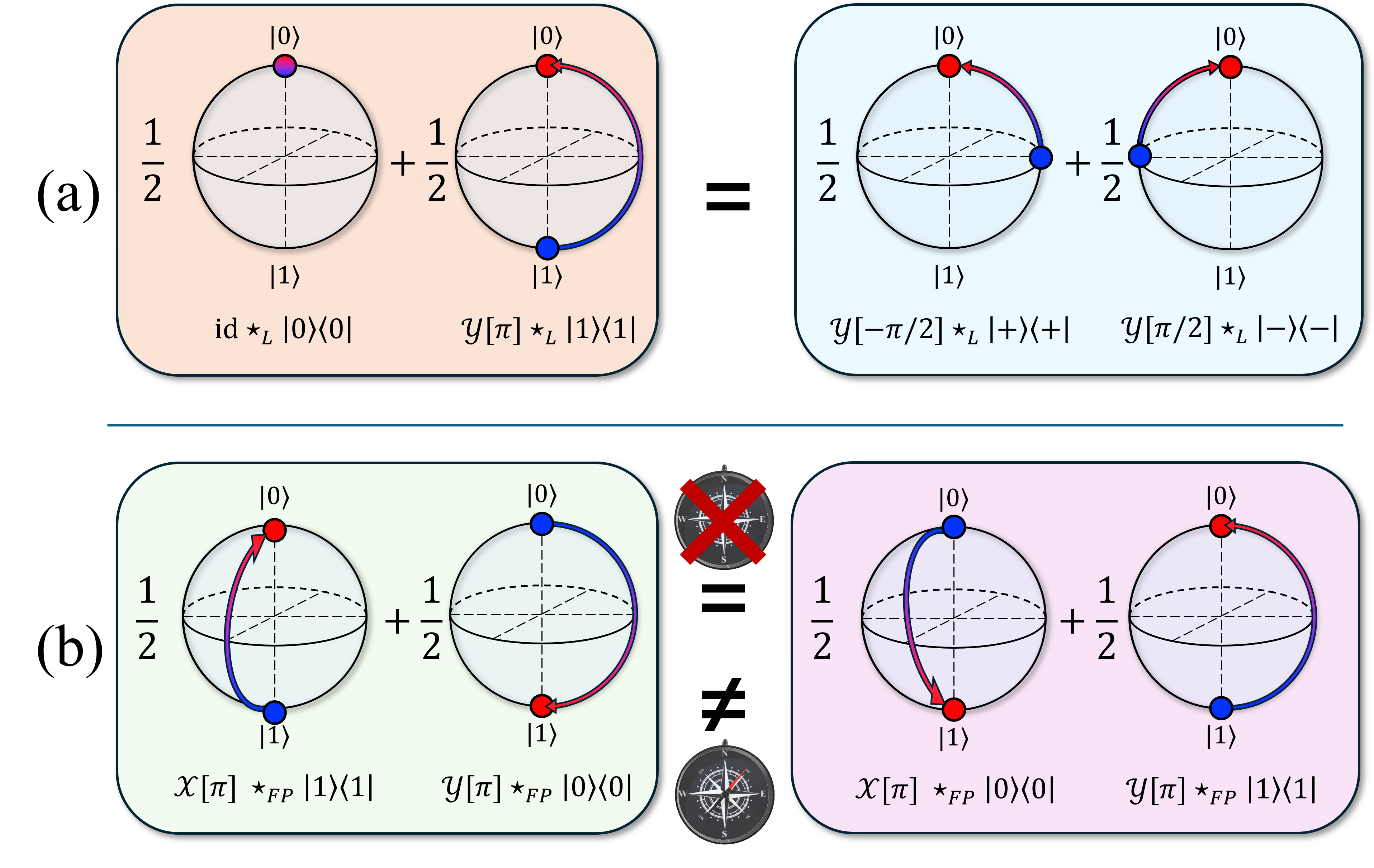}
    \caption{\textit{Examples of interferometrically (in)distinguishable qubit QSOTs} represented by trajectories from blue dot to red dot on the Bloch sphere. (a) (Example \ref{ex:ensemble}) Two vastly different ensembles of qubit dynamics can result in the same QSOT after mixture, becoming indistinguishable under interferometry. (b) (Example \ref{ex:LtoPF}) Certain distinct dynamics indistinguishable under time-reversal symmetry may be distinguished with the help of a temporal reference frame (temporal `compass'). }
    \label{fig:examples}
\end{figure}

It is tempting to conclude that time-reversal symmetry leads to a loss of information accessible through causally agnostic measurements because $\cE \star_{FP}\rho$ is the Hermitian part of $\cE \star_L \rho$ and typically one cannot fully reconstruct an operator from its Hermitian part. We demonstrate that, surprisingly, they are informationally equivalent in the sense that one can recover one from the other for factorizable cases.
\begin{prop} \label{prop:FP=L}
    $\cE_1 \star_{FP} \rho_1 = \cE_2 \star_{FP} \rho_2$ if and only if $\cE_1 \star_L \rho_1 = \cE_2 \star_L \rho_2$ for any states $\rho_1$ and $\rho_2$ on $A$ and channels $\cE_1$ and $\cE_2$ from $A$ to $B$.
\end{prop}
Proof and explicit methods of recovering one expression from the other are given in Sec. \ref{SM:ProPro1} of \cite{SM}. Remarkably, Proposition \ref{prop:FP=L} does not hold for non-factorizable QSOTs with an example given in Example \ref{ex:LtoPF} of End Matter. (See Fig.~\ref{fig:examples} (b).) It implies that one may not be able to access some information in a non-Markovian quantum dynamics without access to a temporal reference frame, i.e., a clock.

\textit{Synchronization as a resource---}
Assume that a composite system $XY$ goes through a parallel dynamics $(\cE_X \otimes \cF_Y,\rho_X \otimes \sigma_Y)$. Compare this with another situation in which $X$ and $Y$ go through two completely unrelated dynamics $(\cE_X,\rho_X)$ and $(\cF_Y,\sigma_Y)$ respectively, maybe in distant regions of spacetime. How are their quantum states over spacetime different? They are given respectively as follows:
\begin{align}
     &(\cE_X \otimes \cF_Y)\star (\rho_X \otimes \sigma_Y), \label{eqn:clockcorr}\\
     \text{and\;\;}&(\cE_X \star \rho_X) \otimes (\cF_Y \star \sigma_Y). \label{eqn:clockcorr2}
\end{align} 
Notably, the equality between \eqref{eqn:clockcorr} and \eqref{eqn:clockcorr2} was previously required as a desirable property named \textit{tensoriality} ~\cite{parzygnat2023axioms} in the context of quantum Bayesian inference. Tensoriality is indeed satisfied by certain QSOT products, e.g., $\star_L$ or $\star_R$, that naturally arise when every dynamics has to share one fixed temporal orientation. However, under time-reversal symmetry, one has to assume $\star=\star_{FP}$ and the equality does not hold in genera. This is because (7) exhibits a correlation between  the dynamics of $X$ and $Y$, such that they always have the same temporal orientation, even if that orientation remains undetermined due to time-reversal symmetry,

whereas \eqref{eqn:clockcorr2} represents complete independence between the dynamics so that they may propagate in different directions in time. (See End Matter for more details.) We call such a correlation between dynamics due to their random but aligned temporal orientations \emph{synchronization}  \footnote{We emphasize that, under broken time-reversal symmetry, synchronization is not possible despite alignment of temporal orientations because no correlation can be formed.}.

Since synchronization is a correlation not simply between local points in spacetime but between two or more dynamics which are already represented with a bipartite QSOT, synchronization is a genuine multipartite correlation in spacetime, which goes beyond the previous classification of bipartite spatiotemporal correlations \cite{song2024causal}.

Even under time-reversal symmetry, by using a reference dynamics synchronized with the dynamics of interest as a \emph{compass} in time, one can access the information about a dynamics in a certain temporal direction.
We present an experimentally feasible example of synchronization and its utilization in the following.

Consider an interferometric protocol for an ensemble of dynamics $(\rho^{(i)},\cE^{(i)})$ from $A$ to $B$ with probability $p_i$, but there is time-reversal symmetry so that one cannot access the QSOT $\sum_i p_i \cE^{(i)}\star_L \rho^{(i)}$ but only $\sum_i p_i \cE^{(i)} \star_{FP} \rho^{(i)}$. However, if we couple this dynamics with a compass initialized in $\ket{0}$ that undergoes the trivial dynamics between $C_A$ and $C_B$ with $C_{A,B}$ being at the same time as $A$ or $B$, then their joint dynamics for each $i$ becomes $(\rho^{(i)}_A\otimes \dyad{0}_{C_A} ,\cE^{(i)} \otimes \id)$ from $AC_A$ to $BC_B$ and its corresponding QSOT is given as:

\begin{equation}
    (\cE^{(i)}_{B|A}\otimes \id_C) \star_{FP} (\rho^{(i)}_A \otimes \dyad{0}_{C_A}),
\end{equation}
where $\id_C$ should be understood as the identity channel from $C_A$ to $C_B$.
The synchronization between $AB$ and $C_AC_B$ dynamics can be utilized as follows. By considering the following intervention unitaries in interferometry:
\begin{gather}
    \tilde{V}_{AC_A} = V_A \otimes \dyad{1}{0}_{C_A} + V^\dag_A \otimes \dyad{0}{1}_{C_A}, \label{eqn:Vtilde}\\
    \tilde{W}_{BC_B} = W_B \otimes \dyad{0}{1}_{C_B} + W^\dag_B \otimes \dyad{1}{0}_{C_B} \label{eqn:Wtilde},
\end{gather}
Note that this argument generalizes to an ensemble of dynamics thus applies to mixed QSOTs too.
This implies that one can recover $\sum_i p_i \cE^{(i)} \star_L \rho^{(i)}$ by repeating this experiment with various $V_A$ and $W_B$ and thus one could distinguish different dynamics that became indistinguishable under time-reversal symmetry, such as the one given in Example \ref{ex:LtoPF} in End Matter, with the help of a compass system.
Such an ability of the compass system originates from its memory effect; it can coherently store quantum information in the form of a bit flip between $\ket{0}$ and $\ket{1}$ and hand it over to the future to break time-reversal symmetry. This clearly demonstrates the physical relevance of synchronization, a multipartite spatiotemporal correlation, and the capability of the QSOT formalism.

This setup is experimentally feasible with current technology as it is just a case of a scattering circuit that requires only one additional qubit undergoing the trivial dynamics of $\id_C$ and the controlled bit-flip-type interventions of \eqref{eqn:Vtilde} and $\eqref{eqn:Wtilde}$, which are standard in qubit platforms. (See \ref{SM:_compass} of \cite{SM}.)

\textit{Conclusion---}We have provided a concrete operational definition of the quantum state over spacetime via causally agnostic measurements implemented using interferometry. In our formalism, the most widely adopted QSOTs naturally emerge as the interference terms in the interferometer’s outcome probabilities. Based on this, mixed quantum states over time can be defined analogously to mixed density matrices, highlighting cases where ensembles of dynamics cannot be distinguished by interferometry.

Our formalism also reveals a clear distinction between the two QSOT products, the left and the FP products, as they capture interferometric inference for dynamics without and with time-reversal symmetry, respectively. Building on this idea, we further demonstrated that two quantum dynamics may exhibit a novel form of spatiotemporal correlation called synchronization, of genuinely multipartite nature, that can be exploited to access temporally asymmetric information. This subtle correlation, which is difficult to capture using conventional formalisms, further underscores the relevance of the QSOT framework. We leave the full characterization and classification of spatiotemporal quantum resources as future work.

\begin{acknowledgments}
    We thank the anonymous referees for constructive inputs and suggestions.
    SHL is supported by the start-up fund and the 2025 Research Fund (1.250007.01) of Ulsan National Institute of Science \& Technology (UNIST), Institute of Information \& Communications Technology Planning \& Evaluation (IITP) Grants (RS-2025-02283189) and National Research Foundation of Korea (RS-2025-25464492). 
    HK is supported by the KIAS Individual Grant No. CG085302 at Korea Institute for Advanced Study and National Research Foundation of Korea (Grants No.~RS-2024-00413957 and No.~RS-2024-00438415) funded by the Korea Government (MSIT).
\end{acknowledgments}

\bibliography{main}

\onecolumngrid
\begin{appendix}
\onecolumngrid
\section*{End Matter}
\twocolumngrid
\twocolumngrid

\section{Examples of indistinguishable dynamics}
The following example demonstrates that two different ensembles of quantum dynamics can lead to the same quantum state over spacetime, which implies excessive information in the conventional channel formalism regarding distinguishability under causally agnostic measurements.
\begin{example} \label{ex:ensemble}
    Two even ensembles of factorizable QSOTs for a qubit system $\qty{\id\star_L\dyad{0}, \cY[\pi]\star_L \dyad{1}}$ and $\qty{\cY[-\pi/2]\star_L \dyad{+}, \cY[\pi/2]\star_L \dyad{-}}$ correspond to the same mixed QSOT
    \begin{align*}
        &\frac{1}{2} \left( \id\star_L\dyad{0} + \cY[\pi]\star_L \dyad{1} \right) \\
        = &\frac{1}{2} \left( \cY[-\pi/2]\star_L \dyad{+} + \cY[\pi/2]\star_L \dyad{-} \right),
    \end{align*}
    with the matrix representation
    $$\frac{1}{2}
    \begin{pmatrix}
        1 &0 &0 &0\\
        0 &0 &1 &0\\
        0 &0 &1 &0\\
        -1&0 &0 &0
    \end{pmatrix},
    $$
    where $\cY [\theta] (\sigma) = \exp(-i \theta Y/2 ) \, \sigma \, \exp(i \theta Y/2 )$ with $Y$ being the Pauli Y operator is the rotation around the $y$-axis of the Bloch sphere by $\theta$.
\end{example}

We also provide another example, where two different mixtures of quantum dynamics that are distinguishable with broken time-reversal symmetry become indistinguishable if we impose time-reversal symmetry. It highlights the importance of temporal reference frames.

\begin{example} \label{ex:LtoPF}
    Consider a qubit system and the Pauli channels $\cX$ and $\cY$. Two QSOTs $\rho^{(1)}_L=\frac12(\cX\star_L\dyad{0}+\cY\star_L\dyad{1})$ and $\rho^{(2)}_L=\frac12(\cY\star_L\dyad{0}+\cX\star_L\dyad{1})$ are distinct as
    $$\rho^{(1)}_L=\frac{1}{2}
    \begin{pmatrix}
        0 &0 &0 &1\\
        0 &1 &0 &0\\
        0 &0 &1 &0\\
        -1 &0 &0 &0
    \end{pmatrix}
    \text{, }
    \rho^{(2)}_L=\frac{1}{2}
    \begin{pmatrix}
        0 &0 &0 &-1\\
        0 &1 &0 &0\\
        0 &0 &1 &0\\
        1 &0 &0 &0
    \end{pmatrix}.
    $$
    
    However, their Hermitian parts, or the corresponding FP-products $\rho^{(1)}_{FP}$ and $\rho^{(2)}_{FP}$ are identical to the classical maximally anti-correlated state
    $$
    \rho^{(1)}_{FP}=\rho^{(2)}_{FP}= 
    \frac{1}{2}
    \begin{pmatrix}
        0 &0 &0 &0\\
        0 &1 &0 &0\\
        0 &0 &1 &0\\
        0 &0 &0 &0
    \end{pmatrix}.
    $$
\end{example}

We present another intriguing example, which is a mixture of two QSOTs whose initial and final states are the same. Even when certain states are unaltered by the given channels, their temporal correlations may be distinct as they can be detected through interferometry. Nevertheless, such QSOTs can still become indistinguishable under time-reversal symmetry.

\begin{example} \label{ex:LtoPF2}
    Consider a qubit system and the rotation $\cZ[\theta]$ around the $z$-axis of the Bloch sphere by $\theta$. Two QSOTs $\rho^{(1)}_L=\frac12(\cZ[\theta]\star_L\dyad{0}+\cZ[-\theta]\star_L\dyad{1})$ and $\rho^{(2)}_L=\frac12(\cZ[-\theta]\star_L\dyad{0}+\cZ[\theta]\star_L\dyad{1})$ are distinct as
    $$\rho^{(1)}_L=\frac{1}{2}
    \begin{pmatrix}
        1 &0 &0 &0\\
        0 &0 &e^{i\theta} &0\\
        0 &e^{-i\theta} &0 &0\\
        0 &0 &0 &1
    \end{pmatrix}
    \text{, }
    \rho^{(2)}_L=\frac{1}{2}
    \begin{pmatrix}
        1 &0 &0 &0\\
        0 &0 &e^{-i\theta} &0\\
        0 &e^{i\theta} &0 &0\\
        0 &0 &0 &1
    \end{pmatrix}.
    $$
    
    However, their Hermitian parts, or the corresponding FP-products $\rho^{(1)}_{FP}$ and $\rho^{(2)}_{FP}$ are identical to
    $$
    \rho^{(1)}_{FP}=\rho^{(2)}_{FP}= \cD[\theta] \star_{FP} \pi =
    \frac{1}{2}
    \begin{pmatrix}
        1 &0 &0 &0\\
        0 &0 &\cos\theta &0\\
        0 &\cos\theta &0 &0\\
        0 &0 &0 &1
    \end{pmatrix},
    $$
    where $\pi=\mds{1}/2$ is the maximally mixed qubit state and $\cD[\theta] = (\cZ[\theta]+\cZ[-\theta])/2$ is the dephasing channel that dampens the off-diagonal elements by the factor of $\cos(\theta/2)$.
\end{example}

\section{Interferometric protocol for dynamics}
Consider the interferometry setting for temporally separated quantum systems. A generic one-step quantum dynamics given as a pair of an initial state $\rho_A$ and a quantum channel $\cE$ from $A$ to $B$. Assume that we are allowed to intervene in this dynamics only at two points, $A$ and $B$, before and after the action of the channel $\cE$. We prepare a reference system $R$ in $\ket{\psi}_R=\alpha_0 \ket{0}_R+ \alpha_1 \ket{1}_R$ and apply intervention unitaries $V_A$ and $W_B$ respectively before and after $\cE$. If $\ket{\psi}_{AE}$ is a purification of $\rho_A$ and $U: AK\to BK$ is a Stinespring dilation (or a unitary extension thereof) of channel $\cE$ so that $\cE(\rho)=\Tr_K[U(\rho_A\otimes \dyad{0}_K)U^\dag]$, then the resultant state of joint system $BEKR$ is
\[
    \alpha_0 U\ket{\psi}_{AE}\ket{00}_{KR} + \alpha_1 W_{B}UV_A\ket{\psi}_{AE}\ket{01}_{KR}.
\]
Finally, by measuring the reference system $R$ in the $\qty{\ket{b_+},\ket{b_-}}$ basis, the outcome probabilities are given as
\begin{equation}
    \Pr(\pm) =  \cS_\pm + 2\Re[\cA_\pm \cI].
\end{equation}
Here, $\cS_\pm = |\alpha_0 \braket{b_\pm}{0}|^2 + |\alpha_1 \braket{b_\pm}{1}|^2$, $\cA_\pm = \alpha_0^*\alpha_1 \braket{0}{b_\pm} \braket{b_\pm}{1}$ and $\cI=\bra{\psi}_{AE}\bra{0}_R U^\dag W_B U V_A \ket{\psi}_{AE}\ket{0}_R$. The interference term $\cI$ can be rewritten as,
$$
\cI = \Tr_A\qty[W_A \Tr_B\qty[U \qty(V_A \Tr_E\qty[\dyad{\psi}_{AE}]\otimes \dyad{0}_R) U^\dag]].
$$
By noting that  $\Tr_E[\dyad{\psi}_{AE}]=\rho_A$ and $\Tr_B[V_A\rho_A\otimes\dyad{0}_R]=\cE(V\rho)$, the interference term $\cI$ simplifies to
\begin{equation} \label{eqn:tempmeasprob}
    \cI = \Tr[W\cE(V\rho)],
\end{equation}
which yields the following, since $\cE(V\rho)=\Tr_A[(V_A\rho_A \otimes \mds{1}_B)J[\cE]]$, by the definition of the left-product $\star_L$,
\begin{equation} 
    \Tr[(V_A\otimes W_B)(\cE\star_L \rho)].
\end{equation}

\section{Comparison with process matrices}
\begin{figure}[!t]
    \centering
    \includegraphics[width=\linewidth]{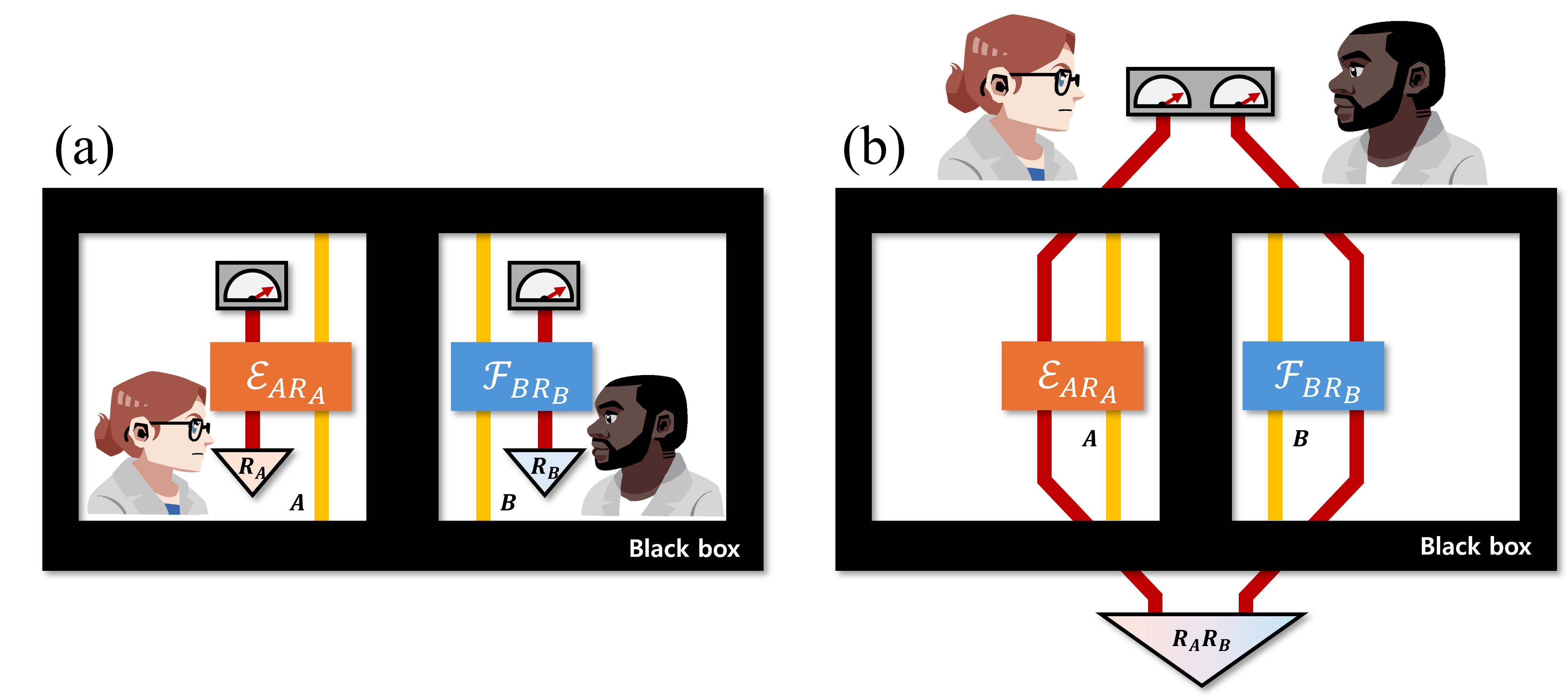}
    \caption{(a) Measurement of process matrix. Alice and Bob, confined in a black box and sharing two quantum systems with an unspecified causal relation, prepare their probe systems $R_A$ and $R_B$ and measure them after they interact with the systems of interest $AB$, individually. (b) However, the same statistics can be obtained by preparing the probe system $R=R_AR_B$ and measuring them after the interaction, before and after the black box. This scheme is almost identical with causally agnostic measurement (Fig.~\ref{fig:agnostic}) except that $A$ and $B$ never interact with the same subset of $R$.}
   \label{fig:process}
\end{figure}

We clarify the difference between causally agnostic measurements and the process matrix formalism introduced in Ref.~\cite{oreshkov2012quantum} in their respective settings, despite some common features. In the \emph{process matrix} framework, two separate experimenters inside a causal black box, Alice and Bob, each operate a laboratory, that receives one quantum system, applies a measurement, and emits another system (See Fig.~\ref{fig:process} (a)).  The most general resource compatible with local quantum mechanics, but \emph{without} assuming any definite causal order between the labs is an operator \(\cW_{AA'BB'}\) acting on the Hilbert space \(AA'BB'\), (un)primed systems being output (input) systems, that provides the probability of Alice and Bob implementing a CP map \(\cM_{AB}:AB\to A'B'\) through the generalized Born rule
\[
\Pr(\cM_{AB})=\Tr[J[\cM_{AB}]\cW_{AA'BB'}].
\]
(We use the basis-independent definition of the process matrix. See Sec. \ref{SM:QSOTPM} of \cite{SM}.) Note that in Ref.~\cite{oreshkov2012quantum}, only product interventions of the form $\cM_{AB}=\cM_A\otimes \cM_B$ were considered, but by sharing an auxiliary system, they can also implement a joint CP map, so we consider a general bipartite CP map $\cM_{AB}$ here. The interferometry considered in the main text is one such example.

The apparent difference between this measurement setting and causally agnostic measurements is that measurements are done within the black box for the former. However, this difference is superficial, since the measurements can be postponed by treating the registers Alice and Bob measure as subsystems $R_A$ and $R_B$ of the probe system $R$ and doing the measurements once the probe system is outside of the black box (See Fig.~\ref{fig:process} (b).). In other words, Alice and Bob need not measure them inside the box.

One of the more fundamental differences arises from the fact that Alice and Bob never interact with the same subsystem of the probe in the process matrix formalism, which requires coordination between Alice and Bob, i.e., a reference frame (e.g., the promise “make sure Alice gets $R_A$ and Bob gets $R_B$”). On the other hand, causally agnostic measurements make no such assumption, so each subsystem can access any part of the probe as long as the interaction is chosen to be invariant under ordering via the controlled unitary operations introduced in Theorem~\ref{thm:CAM}. This results in the possibility that two systems of the same type interact with the probe system in the same fashion, and the agnosticism condition then yields Theorem~\ref{thm:CAM}. (See Sec.~\ref{SM:QSOTPM} of \cite{SM} for a more detailed discussion.) Nevertheless, one could still model the effect of a reference frame in this formalism by including it explicitly as a part of the measured system, as we do in Sec.~\emph{Synchronization as a resource}.

\section{Synchronization under time-reversal symmetry}
In this section, we compare \eqref{eqn:clockcorr} and \eqref{eqn:clockcorr2} to demonstrate their difference and explain alignment of temporal orientations. Note that \eqref{eqn:clockcorr} equals to (recall that the right-product $\star_R$ represents the dynamics in the direction opposite to the left-product $\star_L$.)
\[
   \frac{1}{2} [(\cE_X\star_L \rho_X)\otimes (\cF_Y\star_L \sigma_Y) + (\cE_X\star_R \rho_X)\otimes (\cF_Y\star_R \sigma_Y)],
\]
whereas \eqref{eqn:clockcorr2} equals to 

\begin{gather*}
    \frac{1}{4} [(\cE_X\star_L \rho_X)\otimes (\cF_Y\star_L \sigma_Y) + (\cE_X\star_L \rho_X)\otimes (\cF_Y\star_R \sigma_Y)\\
    +(\cE_X\star_R \rho_X)\otimes (\cF_Y\star_L \sigma_Y) + (\cE_X\star_R \rho_X)\otimes (\cF_Y\star_R \sigma_Y)].
\end{gather*}

Similarly to the difference between the Bell state $\ket{\uparrow\uparrow}+\ket{\downarrow\downarrow}$ and a product state $(\ket{\uparrow}+\ket{\downarrow})\otimes(\ket{\uparrow}+\ket{\downarrow})$, \eqref{eqn:clockcorr} exhibits random but aligned orientations but \eqref{eqn:clockcorr2} shows no correlation at all.

\section{Pure, mixed, and (non-)factorizable QSOTs}
    Formally speaking, a bipartite QSOST $\rho_{AB}$ on systems $AB$ is \emph{factorizable} in the $A \to B$ direction with respect to a QSOT product $\star$ if it is compatible with time-like separated $A$ and $B$ and if there exists a quantum channel $\cE$ from $A$ to $B$ such that $\rho_{AB}=\cE\star\rho_A$ for $\rho_A=\Tr_B\rho_{AB}$. Note that factorizability depends on the choice of $\star$.
    
    There is a certain amount of analogy one can draw between pure quantum states and factorizable QSOTs, but they are not completely analogous. Especially, if one defined a pure (one-way) QSOT as an extremal point in the set ${\tt QSOT}_{A\to B}$ of all QSOTs from system $A$ to $B$, defined as the collection of all operators obtained from partial trace of factorizable QSOTs, i.e.,
    \begin{equation}
        {\tt QSOT}_{A\to B} := \qty{\Tr_{E_AE_B}[\cE_{AE_A\to BE_B}\star \rho_{AE_A}]|\cE,\rho},
    \end{equation}
    for all CPTP maps $\cE:AE_A\to BE_B$ and density operators $\rho_{AE_A}$, then the pure states and the factorizable QSOTs are not identical. For example, $\cE \star \rho$ with mixed $\rho=\sum_i p_i \dyad{\psi_i}$ can be decomposed into $\sum_i p_i \cE \star \dyad{\psi_i}$. Thus, the notions of a mixed QSOT, defined as a non-pure QSOT, and a non-factorizable QSOT are also distinct. Nevertheless, they still share significant similarity. For example, similarly to how one can purify a mixed state into a pure state in a larger space, one can ``dilate'' a non-factorizable QSOT into a factorizable QSOT in a larger space. For example, consider a non-factorizable QSOT $\sum_i p_i \cF_i \star \rho_i$. This is a marginal state of a QSOT defined on a larger space given as $\qty(\sum_i \cF_i \otimes \dyad{i} \; \cdot \; \dyad{i}) \star \qty(\sum_j p_j \rho_j \otimes \dyad{j}).$ The complete mathematical characterization of the set ${\tt QSOT}_{A\to B}$ and the inclusion relation between the classes pure, mixed, (non-)factorizable QSOT are left as an open problem.
\end{appendix}

\clearpage
\onecolumngrid
\begin{center}
  {\bf\large Supplemental Material for}\\[0.5ex]
  {\bf\large ``Probing Quantum States Over Spacetime Through Interferometry''}\\
\end{center}
\section{Useful QSOTs}
Hereby, we list the matrix representation of some useful qubit QSOTs. $\cX, \cY, \cZ$ stand for Pauli channels. For the initial state $\ket{0}$ cases,
$$\id\star_L \dyad{0}=
\begin{pmatrix}
1 & 0 & 0 & 0 \\
0 & 0 & 1 & 0 \\
0 & 0 & 0 & 0 \\
0 & 0 & 0 & 0 \\
\end{pmatrix}, \;
\cZ\star_L \dyad{0}=
\begin{pmatrix}
1 & 0 & 0 & 0 \\
0 & 0 & -1 & 0 \\
0 & 0 & 0 & 0 \\
0 & 0 & 0 & 0 \\
\end{pmatrix}, \;
\cX\star_L \dyad{0}=
\begin{pmatrix}
0 & 0 & 0 & 1 \\
0 & 1 & 0 & 0 \\
0 & 0 & 0 & 0 \\
0 & 0 & 0 & 0 \\
\end{pmatrix}, \;
\cY\star_L \dyad{0}=
\begin{pmatrix}
0 & 0 & 0 & -1 \\
0 & 1 & 0 & 0 \\
0 & 0 & 0 & 0 \\
0 & 0 & 0 & 0 \\
\end{pmatrix}.$$

Similarly, for the initial state $\ket{1}$ cases,
$$\id\star_L \dyad{1}=
\begin{pmatrix}
0 & 0 & 0 & 0 \\
0 & 0 & 0 & 0 \\
0 & 1 & 0 & 0 \\
0 & 0 & 0 & 1 \\
\end{pmatrix}, \;
\cZ\star_L \dyad{1}=
\begin{pmatrix}
0 & 0 & 0 & 0 \\
0 & 0 & 0 & 0 \\
0 & -1 & 0 & 0 \\
0 & 0 & 0 & 1 \\
\end{pmatrix}, \;
\cX\star_L \dyad{1}=
\begin{pmatrix}
0 & 0 & 0 & 0 \\
0 & 0 & 0 & 0 \\
0 & 0 & 1 & 0 \\
1 & 0 & 0 & 0 \\
\end{pmatrix}, \;
\cY\star_L \dyad{1}=
\begin{pmatrix}
0 & 0 & 0 & 0 \\
0 & 0 & 0 & 0 \\
0 & 0 & 1 & 0 \\
-1 & 0 & 0 & 0 \\
\end{pmatrix}.$$

The following formulas for unphysical initial states are also useful for calculation. 
$$\id\star_L \dyad{0}{1}=
\begin{pmatrix}
0 & 1 & 0 & 0 \\
0 & 0 & 0 & 1 \\
0 & 0 & 0 & 0 \\
0 & 0 & 0 & 0 \\
\end{pmatrix}, \;
\cZ\star_L \dyad{0}{1}=
\begin{pmatrix}
0 & -1 & 0 & 0 \\
0 & 0 & 0 & 1 \\
0 & 0 & 0 & 0 \\
0 & 0 & 0 & 0 \\
\end{pmatrix}, \;
\cX\star_L \dyad{0}{1}=
\begin{pmatrix}
0 & 0 & 1 & 0 \\
1 & 0 & 0 & 0 \\
0 & 0 & 0 & 0 \\
0 & 0 & 0 & 0 \\
\end{pmatrix}, \;
\cY\star_L \dyad{0}{1}=
\begin{pmatrix}
0 & 0 & 1 & 0 \\
-1 & 0 & 0 & 0 \\
0 & 0 & 0 & 0 \\
0 & 0 & 0 & 0 \\
\end{pmatrix}.$$

$$\id\star_L \dyad{1}{0}=
\begin{pmatrix}
0 & 0 & 0 & 0 \\
0 & 0 & 0 & 0 \\
1 & 0 & 0 & 0 \\
0 & 0 & 1 & 0 \\
\end{pmatrix}, \;
\cZ\star_L \dyad{1}{0}=
\begin{pmatrix}
0 & 0 & 0 & 0 \\
0 & 0 & 0 & 0 \\
1 & 0 & 0 & 0 \\
0 & 0 & -1 & 0 \\
\end{pmatrix}, \;
\cX\star_L \dyad{1}{0}=
\begin{pmatrix}
0 & 0 & 0 & 0 \\
0 & 0 & 0 & 0 \\
0 & 0 & 0 & 1 \\
0 & 1 & 0 & 0 \\
\end{pmatrix}, \;
\cY\star_L \dyad{1}{0}=
\begin{pmatrix}
0 & 0 & 0 & 0 \\
0 & 0 & 0 & 0 \\
0 & 0 & 0 & -1 \\
0 & 1 & 0 & 0 \\
\end{pmatrix}.$$
The $\star_R$ and $\star_{FP}$ QSOTs can be obtained by calculating the adjoint matrix or the Hermitian part of the corresponding QSOT.

\section{Proof of Proposition 1} \label{SM:ProPro1}
\begin{proof}
    $(\implies)$ Since $\cE_i \star_{FP} \rho_i$ for $i=1,2$ is the Hermitian part of $\cE_i\star_L \rho_i$ as long as $\cE_i$ is Hermitian-preserving and $\rho_i$ is Hermitian, the if part obviously follows. To show the only if part, suppose that $\cE_1 \star_{FP} \rho_1 = \cE_2 \star_{FP} \rho_2$. From the marginality condition, we can easily see that $\rho_1 = \rho_2 = \rho$. Then, we have the condition
$$
\begin{aligned}
 &\cE_1 \star_{FP} \rho - \cE_2 \star_{FP} \rho = 0 \\
 \Leftrightarrow& \left( J[\cE_1] - J[\cE_2] \right) (\rho \otimes \mds{1}) + (\rho \otimes \mds{1})  \left( J[\cE_1] - J[\cE_2] \right) = 0 \\
 \Leftrightarrow& \left( J[\cE_1] - J[\cE_2] \right) (\rho \otimes \mds{1}) = 0 = (\rho \otimes \mds{1}) \left( J[\cE_1] - J[\cE_2] \right),
\end{aligned}
$$
where we used the fact that for a hermitian matrix $A$ and positive semidefinite matrix $B \geq 0$, $AB + BA =0$ if and only if $AB = 0 = BA$ ($\because$ For each eigenstate of $B$, such that $B \ket{\mu} = b_\mu \ket{\mu}$, we have $b_\mu A \ket{\mu} + B A \ket{\mu} = 0 \Leftrightarrow B (A \ket{\mu}) = - b_\mu (A \ket{\mu})$. As $B$ is positive semi-definite, the only possible solutions of this eigenvalue equation are $b_\mu = 0$ or $A \ket{\mu} = 0$. For both cases, we have $b_\mu A \ket{\mu}\bra{\mu} = 0 = b_\mu \ket{\mu}\bra{\mu} A$ so that summing over $\mu$ leads to $BA = 0 = AB$.). Hence, the last condition directly implies that  $\cE_1 \star_{L} \rho_1  = \cE_2\star_{L} \rho_2$ and $\cE_1 \star_{R} \rho_1 = \cE_2\star_{R} \rho_2$ by the definitions of QSOTs.    

$(\impliedby)$ The converse follows immediately from the observation that $\cE_i\star_{FP}\rho_i$ is the Hermitian part of $\cE_i\star_L\rho_i$ for $i=1,2$, so $\cE_i\star_{FP}\rho_i=[\cE_i\star_L\rho +(\cE_i\star_L\rho)^\dag]/2$ is uniquely determined once $\cE_i\star_L\rho_i$ is specified.

More explicitly, one can recover $\cE \star_{L} \rho$ from $\cE \star_{FP} \rho$ as follows. Assume that $\rho=\sum_i \lambda_i P_i$ is the spectral decomposition of $\rho$ with $\lambda_i$ and $P_i$ being its eigenvalues and projectors onto the corresponding eigenspace. Also assume that $\Pi$ and $\Pi^\perp$ are projectors onto the support and kernel of $\rho$, respectively. Then note that if we define functions
\[
    \Theta(X):= \frac{1}{2} \{\rho,X\},
\]
and
\[
    \Theta^{-1}(X):=\sum_{\substack{i,j:\\ \lambda_i+\lambda_j\neq0}} \frac{2}{\lambda_i+\lambda_j} P_i X P_j,
\]
they are inverse to each other whenever $\rho$ is full-rank, and if $\rho$ is rank-deficient,
\[
    \Theta \circ \Theta^{-1} (X) = \Theta^{-1}\circ \Theta (X) = X - \Pi^\perp X \Pi^\perp.
\]
Since the FP-product can be expressed as $\cE\star_{FP}\rho=(\Theta\otimes \id_B) J[\cE]$ in terms of $\Theta$, one can recover $J'[\cE]:=J[\cE]- (\Pi^\perp \otimes \mds{1}_B) J[\cE] (\Pi^\perp \otimes \mds{1}_B)$ by applying $\Theta^{-1}\otimes \id_B$ to $\cE\star_{FP}\rho$. 
This directly leads to the reconstruction of ${\cal E} \star_L \rho$ by noting that $(\rho_A \otimes \mds{1}_B) J'[\cE]=(\rho_A \otimes \mds{1}_B) J[\cE]=\cE\star_L\rho$, where the first equality holds as $\rho\Pi^\perp=0$.

\end{proof}

\section{Derivation of interference term under time-reversal symmetry} \label{SM:DervIntTRev}
In this section, we derive the expression of the interference term associated with a dynamics under time-reversal symmetry. Consider a dynamics $(\rho, \cE)$ from $A$ to $B$. Suppose that $\cE(\rho)=\Tr_{E_B}[\cU(\rho_A\otimes \tau_{E_A})]$ is a dilation of $\cE$ with some auxiliary systems $E_A$ and $E_B$ at two different times and a pure state $\tau_{E_A}$. The unitary evolution $\cU$ from $AE_A$ to $BE_B$ given as $\cU(X)=UXU^\dag$ for some unitary operator $U:AE_A\to BE_B$ is reversible even when $\cE$ itself is not. We have already established that the interference term $\cI_{\text{fwd}}$ associated with the forward dynamics is given as $\Tr[W\cE(V\rho)]$ in the main text. The time-reversed dynamics of the dilated dynamics $(\rho_A\otimes\tau_{E_A},\cU)$ is $(\cU(\rho_A\otimes \tau_{E_A}),\cU^\dag)$. The associated interference term $\cI_{\text{rev}}$ is given as
\begin{align*}
    \cI_{\text{rev}}=& \Tr[(V_A\otimes \mds{1}_{E_A})\cU^\dag\qty((W_B\otimes\mds{1}_{E_B})\cU(\rho_A\otimes \tau_{E_A}))]\\
    =&\Tr[(V_A\otimes \mds{1}_{E_A})U^\dag\qty((W_B\otimes\mds{1}_{E_B})U(\rho_A\otimes \tau_{E_A})U^\dag)U]\\
    =&\Tr[(W_B\otimes\mds{1}_{E_B})U(\rho_AV_A\otimes \tau_{E_A})U^\dag]\\
    =&\Tr[W\Tr_{E_B}[U(\rho_AV_A\otimes \tau_{E_A})U^\dag]]\\
    =&\Tr[W\cE(\rho V)].
\end{align*}
Here, the first equality is the result of the same argument for deriving the interference term given in the main text applied to the dynamics $(\cU(\rho_A\otimes\tau_{E_A})$
The second equality follows from $\cU(X)=UXU^\dag$ and $\cU^\dag(X)=U^\dag X U$, and the third and fourth equality hold because of the cyclicity of trace and $U^\dag U = \mds{1}$. The fifth equality is by $\Tr=\Tr_B\otimes \Tr_{E_B}$ and the sixth holds because $\cE(X)=\Tr_{E_B}[U(X\otimes \tau_{E_A})U^\dag]$. Now, since the interference term is linear with respect to the statistical mixture, we have that the total inference term $\cI$ is an even mixture of $\cI_{\text{fwd}}$ and $\cI_{\text{rev}}$, which results in
$$\cI = \frac{1}{2}\Tr[W\cE(V\rho)+W\cE(\rho V)].$$

\section{Characterizing interferometry} \label{SM:CharInt}
In a causally agonistic measurement of two systems $A$ and $B$, the probe system $R$ interacts with $A$ and $B$ in an unspecified order and in an unorchestrated fashion. The description of that interaction specified by the Hamiltonians $H_{AR}$ and $H_{BR}$ and interaction times $t_A$ and $t_B$ should be independent of the order for arbitrary $t_A$ and $t_B$, as long as they are independently picked from the set of admissible interaction Hamiltonians $\fI$. It amounts to the following in terms of quantum channels.
\begin{enumerate}
    \item (No-orchestration) The class of causally agnostic measurement schemes is characterized only by the set of admissible Hamiltonians $\fI$. It means that the type and the duration of interaction between the probe and the system of interest can be chosen independently and spontaneously at each region in spacetime as long as the interaction Hamiltonian is picked from the admissible set $\fI$.
    \item (Commutativity) For any two systems $X$ and $Y$, as long as $H_{XR}$ and $H_{YR}$ are admissible, i.e. in $\fI$, the interaction unitary channels $\cU_{AH}^{t_X}(\sigma)=\exp(-iH_{XR}t_X)\sigma\exp(iH_{XR}t_X)$ and $\cU^{t_Y}_{YH}(\sigma)=\exp(-iH_{YR}t_Y)\sigma\exp(iH_{YR}t_Y)$ commute for any positive interaction time $t_X$ and $t_Y$. In other words,
    $$(\cU_{XR}^{t_X}\otimes \id_Y) (\id_X \otimes \cU_{YR}^{t_Y})=(\cU_{XR}^{t_X}\otimes \id_Y) (\id_X \otimes \cU_{YR}^{t_Y}).$$
\end{enumerate}

However, by considering infinitesimal interaction times $t_X$ and $t_Y$, we have $\cU_{AH}^{t_X}(\sigma)= \sigma -i t_X [H_{XR},\sigma] + O(\epsilon^2)$ and similarly for $\cU^{t_Y}_{YH}(\sigma)$, so that one has the following equality for any operator $\sigma$ on $XYR$,
\[
    [H_{XR}\otimes \mds{1}_Y,[H_{YR}\otimes \mds{1}_X,\sigma]]=[H_{YR}\otimes \mds{1}_X,[H_{XR}\otimes \mds{1}_Y,\sigma]],
\]
which is equivalent to
\[
    [[H_{XR}\otimes \mds{1}_Y,H_{YR}\otimes \mds{1}_X],\sigma]=0.
\]
Since it holds for arbitrary $\sigma$, from this we get $[H_{XR}\otimes \mds{1}_Y,H_{YR}\otimes \mds{1}_X]=c \mds{1}_{XYR}$. However, in the case of finite-dimensional systems, the left-hand side is traceless, hence $c=0$, and we conclude that
\[
    [H_{XR}\otimes \mds{1}_Y,H_{YR}\otimes \mds{1}_X]=0.
\]

Because of the well-known commuting-subalgebra structure theorem~\cite{bravyi_vyalyi_2005,wedderburn_1908,kadison_ringrose_1997,jacobson_1989}, by considering two commuting subalgebras $\mathbf{Alg}\qty{\bra{i}_X H_{XR} \ket{j}_X|i,j}$ and $\mathbf{Alg}\qty{\bra{i}_Y H_{YR} \ket{j}_Y|i,j}$ ($\mathbf{Alg}\; S$ stands for the algebra generated by the set $S$ and $\qty{\ket{i}_X}$ is the computational basis of $\cH_X$ and similarly for $Y$.) of the operator algebra on $\cH_R$, it follows that the Hilbert space $\cH_R$ of the probe system $R$ decomposes into $\bigoplus_i \cH_{R,i}^L \otimes \cH_{R,i}^R$ and $H_{XR}$ and $H_{YR}$ also decompose into
\begin{equation} \label{eqn:SM_HamilComm}
    H_{XR} = \bigoplus_i H_{XR^L_i} \otimes \mds{1}_{R^R_i} \text{ and } H_{YR} = \bigoplus_i H_{YR^R_i} \otimes \mds{1}_{R^R_i},
\end{equation}
where $H_{XR^L_i}$ is a Hamiltonian defined on $\cH_X\otimes \cH^L_{R,i}$ and similarly $H_{YR^R_i}$ on $\cH_X\otimes \cH^R_{R,i}$ for each $i$.

However, by the no-orchestration condition, for any system $X$ and any Hamiltonian $H_{XR}$ in the admissible set $\fI$, another system $X'$ of the same type can go through an interaction governed by the same Hamiltonian $H_{XR}$ (only with the change of label $X\to X'$). It means that the decomposition requirement of \eqref{eqn:SM_HamilComm} has to be applied to two copies of $H_{XR}$, but it is possible only when $\cH_R=\bigoplus_i \cH_{R,i}$ without the tensor decomposition of each $\cH_{R,i}$ into the $L$ and $R$ parts, and when $H_{XR}=\bigoplus_i H_X^{(i)}\otimes \mds{1}_{R_i}$ for some Hamiltonian $H_X^{(i)}$ on $\cH_X$ with $\mds{1}_{R_i}$ being the identity on $\cH_{R,i}$ for each $i$. This amounts to saying that unitary operators induced by $H_{XR}$ are controlled unitary operators in the form of
\begin{equation}
    U_{XR} = \bigoplus_i U^{(i)}_X \otimes \mds{1}_{R_i}.    
\end{equation}
Thus, for arbitrary causally agnostic measurement scheme involving systems $A_1,A_2,\dots,A_n$ as considered in Theorem 1, whenever the initial state of the probe system $R$ is given as $\ket{\psi}_R=\sum_i \alpha_i \ket{\psi_i}_R$ where $\ket{\psi_i}_R\in \cH_{R,i}$, the whole system is in the superposition of pure states of $A_1A_2\cdots A_n$ with unitary operators $V^{(i)}_{A_j}$ acted upon them with amplitude $\alpha_i$. This precisely corresponds to an interferometry with arms labeled with the index $i$.

\section{Quantum states over spacetime and process matrices} \label{SM:QSOTPM}

In this Section, we compare the process matrix formalism developed in Refs. ~\cite{oreshkov2012quantum,araujo2014quantumcontrol,procopio2015suporder,araujo2015witness,oreshkov2016causal} with the quantum state over spacetime formalism developed in our work and clarify the connection between the formalisms. A \textit{process matrix} $\cW$~\cite{oreshkov2012quantum,araujo2014quantumcontrol,procopio2015suporder,araujo2015witness,oreshkov2016causal} associated with a bipartite system $A$ and $B$ with an unspecified causal relation possessed by Alice and Bob, respectively, is a matrix on $AA'BB'$ that yields the probability of Alice and Bob implementing a probabilistic process represented by a trace-nonincreasing CP map $\cM_{AB}:AB \to A'B'$ through the following generalized Born rule:
\begin{equation}
    \Pr(\cM_{AB})=\Tr[J[\cM_{AB}] \cW_{AA'BB'}].
\end{equation}
Here, we use the basis-independent definition by using the \jami isomorphism of $\cM_{AB}$ given as $J[\cM_{AB}]=(\id_{AB}\otimes \cM_{AB})(\swap_{AB})$ (Here, $\swap_{AB}=\sum_{ii'jj'} \dyad{ii'}{jj'}_{AB}\otimes\dyad{jj'}{ii'}_{AB\text{ or } A'B'}$ is the swap operator between $AB$ and either its unnamed copy isomorphic to $AB$ or $A'B'$ depending on the place it appears.) because of mathematical simplicity, but note that it is equivalent to the original definition $\cW_{og}$ of Ref.~\cite{oreshkov2012quantum} up to partial transpositions, i.e. $\cW=\cW^{T_{AB}}_{og}$.
While Ref.~\cite{oreshkov2012quantum} focused on product interventions of the form $\cM_{AB}=\cM_A\otimes \cM_B$, here we consider a more general form of bipartite CP map $\cM_{AB}$, which includes joint CP map implementable by sharing an auxiliary system. Note that we always have the normalization condition $\Tr[J[\cM_{AB}]\cW]=1$ for any CPTP map $\cM_{AB}$ and the Hermiticity condition $\cW=\cW^\dag$.

\begin{figure}
    \centering
    \includegraphics[width=0.8\linewidth]{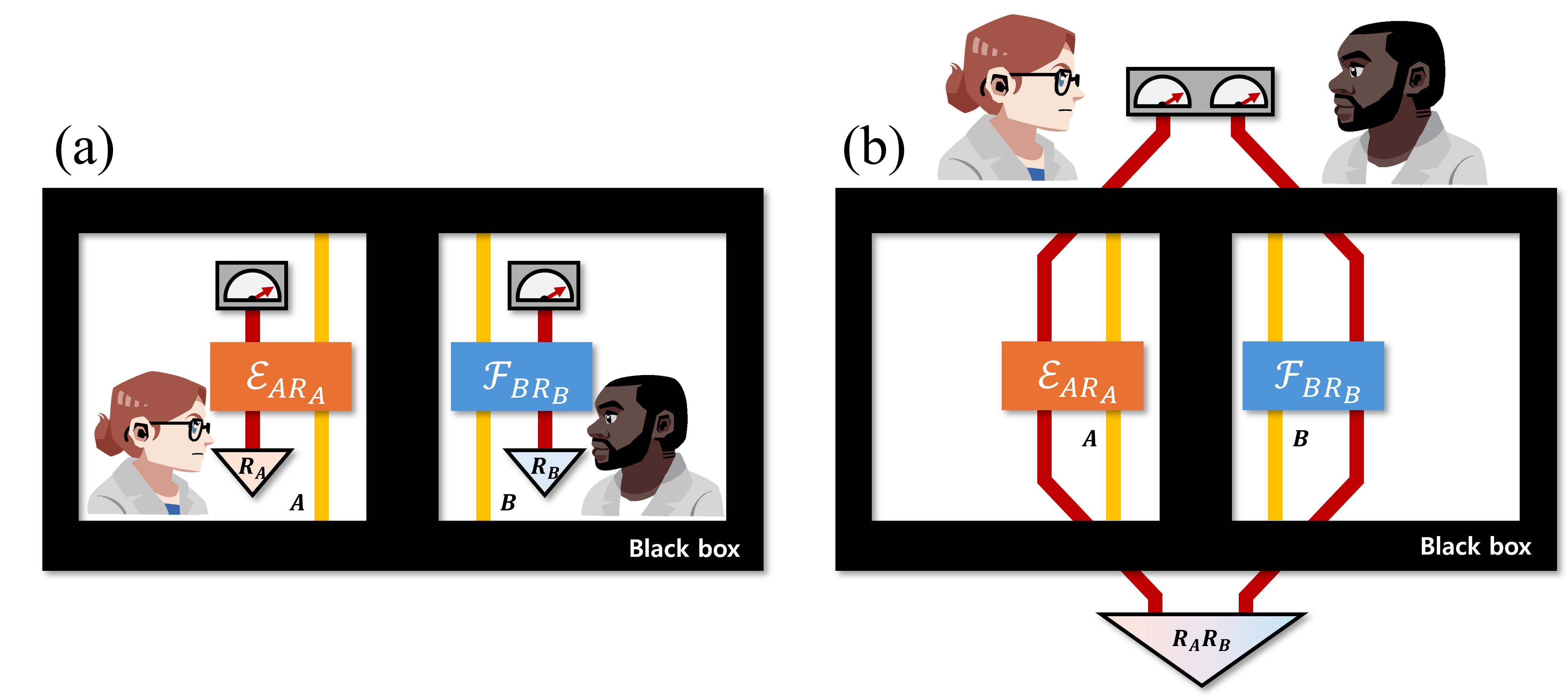}
    \caption{(Fig. \ref{fig:process} of the main text.) A process matrix is an operator that yields the probability of Alice and Bob measuring a specific outcome when they are allowed to have an arbitrary causal relation. While the original setting considered in Ref. \cite{oreshkov2012quantum} can be described as (a) in which both Alice and Bob have uncorrelated probe states and their own independent measurement devices, this setting can be generalized to that of (b), where their probe systems are prepared outside of the causal black box, and the probe systems are jointly measured outside of the box.}
    \label{fig:processcomp2}
\end{figure}

Let us consider the \textit{first-order approximation} of the process matrix $\cW$ when Alice and Bob implement a joint weak measurement close to the trivial measurement, i.e., $\cM_{AB}=p\;\id_{AB}$ with $0\leq p \leq 1$. This can be modeled with a CP map $\cM_{AB}(X)=p(\mds{1}_{AB}-K_{AB}/2)X(\mds{1}_{AB}-K^\dag_{AB}/2)$ with some small operator $K_{AB}$ such that $\norm{K_{AB}}\leq \epsilon$. (The more general case of higher Choi rank can always be decomposed into this simplest Choi rank-1 case, hence we will only consider this case. Higher Choi rank CP maps can always be understood as a coarse-graining of Choi rank-1 CP maps.) The corresponding POVM element $M_{AB} := p|\mds{1}_{AB}-K_{AB}/2|^2 $ is assumed to satisfy $M_{AB}\leq \mds{1}_{AB}$. Then the probability of obtaining this outcome is given as
\begin{equation}\label{eqn:process_matrix_linear_approx}
\begin{aligned}
        \Pr(\cM_{AB}) &= \Tr \left[ \left(  \id_{AB} \otimes \cM_{AB}\right) \left( \swap_{AB} \right) \cW_{AA'BB'} \right]\\
        &= p \Tr \left[ \swap_{AB}  \cW_{AA'BB'} - (\mds{1}_{AB} \otimes K_{A'B'}/2)\swap_{AB}  \cW_{AA'BB'} - \swap_{AB} (\mds{1}_{AB} \otimes K^\dagger_{AB}/2)  \cW_{AA'BB'} \right]\\
        &\quad + p \Tr \left[ (\mds{1}_{AB} \otimes K_{A'B'}/2)\swap_{AB} (\mds{1}_{AB} \otimes K^\dagger_{A'B'}/2) \cW_{AA'BB'} \right]\\
        &= p \left( \Tr \left[ \swap_{AB}  \cW_{AA'BB'} \right] - \frac{1}{2} \Tr \left[(\mds{1}_{AB} \otimes K_{A'B'})\swap_{AB}  \cW_{AA'BB'} +  (\mds{1}_{AB} \otimes K^\dagger_{AB}) ( \swap_{AB} \cW_{AA'BB'} )^\dagger  \right] \right)\\
        &\quad + O(\epsilon^2)\\
        &= p \Re\qty[\Tr[\qty(\mds{1}_{A'B'}-K_{A'B'}) \cW^{(1)}]] + O(\epsilon^2),    
\end{aligned}
\end{equation}
where we define
\begin{equation} \label{eqn:cW1}
    \cW^{(1)}=\Tr_{AB}[\swap_{AB} \cW_{AA'BB'}],
\end{equation}
with $\Tr_{AB}$ being the partial trace over unprimed system $AB$, and the second-order term containing both $K_{A'B'}$ and $K_{A'B'}^\dagger$ is suppressed as $O(\epsilon^2)$. As $\cW^{(1)}$ is an operator on the reduced system $A'B'$ after tracing out the reference system $AB$, it contains less amount of information than $\cW_{AA'BB'}$. Nevertheless, Eq.~\eqref{eqn:process_matrix_linear_approx} shows that $\cW^{(1)}$ fully captures the linear contribution of the weak measurement effects (i.e., terms including only $K_{A'B'}$ or $K_{A'B'}^\dagger$), which justify it as a proper first-order approximation of the process matrix. We also note that the non-Hermiticity of $\cW^{(1)}$ is the artifact of time-reversal asymmetry of the weak measurement process given above. Note that $\cW^{(1)}$ is normalized since $\Tr\cW^{(1)}=\Tr[\swap_{AB}\cW_{AA'BB'}]=\Tr[J[\id_{AB\to A'B'}]\cW_{AA'BB'}]=1$ where $\id_{AB\to A'B'}$ is the identity map and hence a CPTP map, and $\cW$ is already normalized for CPTP maps.

It turns out that the first-order term $\cW^{(1)}$ of the process matrix has broad applicability outside of the weak measurement scheme, as it could describe any scenario in which the first-order expansion of $\cM_{AB}$ is relevant.
In particular, we show that for the interferometry setting considered in the main text, the second-order term in Eq.~\eqref{eqn:process_matrix_linear_approx} entirely vanishes so that $\cW^{(1)}$ contains all the information of the interferometer outcomes. This can be explicitly derived by taking $\cM^\pm_{AB}(X)=(\mds{1}_{AB} \pm V_{A'}\otimes W_{B;}) X (\mds{1}_{AB} \pm V^\dag_{A'}\otimes W^\dag_{B'})/4$, i.e., $p=1/4$ and $K_{A'B'}/2 = \mp V_{A'} \otimes W_{B'}$. This can be shown by assuming the superposed initial state $\ket{\psi}_R=(\ket{0}_R+\ket{1}_R)/\sqrt{2}$, which describes the equal superposition of two paths in an interferometer. Applying conditional $V_A$ and $W_B$ operations on $A$ and $B$ respectively followed by the measurement of the probe system $R$ with respect to the $\qty{\ket{+},\ket{-}}$ basis yields the expression of $\cM_{AB}$ above. We then note that for a unitary operator $V_A \otimes W_B$, the second-order term in Eq.~\eqref{eqn:process_matrix_linear_approx} becomes $\Tr \left[ (\mds{1}_{AB} \otimes V_{A'} \otimes W_{B'})\swap_{AB} (\mds{1}_{AB} \otimes V_{A'}^\dagger \otimes W_{B'}^\dagger) \cW_{AA'BB'} \right] =   \Tr[J[\cU_{AB}] \cW_{AA'BB'}] = \Pr(\cU_{AB})=1$, where $\cU_{AB}(X) = (V_A \otimes W_B) X (V_{A} \otimes W_B)^\dagger$ is a unitary, and thus a CPTP, map. This leads to the probabilities for the two outcomes as
\begin{equation}
    \Pr(\pm) = \Pr(\cM_{AB}^\pm)=\frac{1\pm \Re[\Tr[(V_A\otimes W_B)\cW^{(1)}]]}{2}.
\end{equation}
According to the definition of the quantum state over spacetime proposed in the main text, we conclude that $\cW^{(1)}$ is equivalent to the quantum state over spacetime $\rho_{AB}$ of $A$ and $B$.

In conclusion, one can access the quantum state over spacetime with various methods, as long as they can probe the first-order term in the expansion of the process matrix of the given quantum systems around the trivial measurement. In some sense, the quantum state over spacetime formalism is not a generalization but rather a specialization of the process matrix formalism; however, this is not a weakness. The situation is similar to how the density-operator formalism provides weaker distinguishability compared to the pure-state ensemble formalism, yet the former is more relevant in operational quantum mechanics, since there are no measurement methods that can distinguish different ensembles yielding the same density operator, and all operationally accessible information is captured by the density operator itself. Similarly, if the given measurement setting can only distinguish the immediate response of the quantum systems in spacetime in the form of the first-order term in the expansion, then using the more succinct formalism of quantum states over spacetime is more appropriate than using the full process matrix formalism. (For example, see Sec. \ref{SM:Markov} to see how the quantum state over spacetime formalism is sufficient for defining quantum non-Markovianity without calculating the whole process matrix.)

\section{Derivation of compass interference term} \label{SM:CompInt}
In this section, we derive that one can access the left-product interference term with the help of a compass system whose dynamics is synchronized with that of the system of interest.
Suppose that a dynamics associated with two composite systems $AC_A$ and $BC_B$ at two different times is given as an FP product

\begin{figure}
    \centering
    \includegraphics[width=0.5\linewidth]{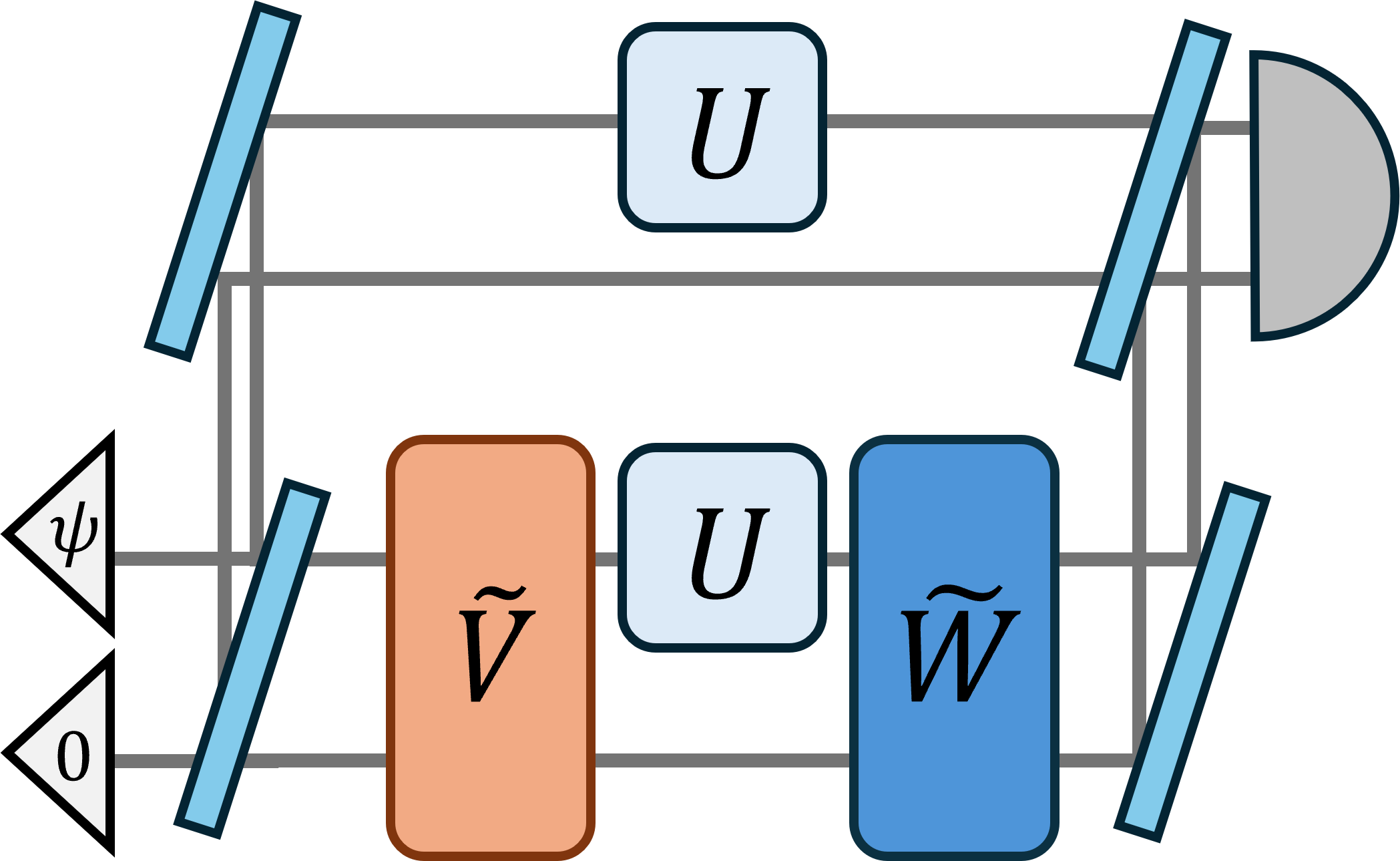}
    \caption{Interferometric diagram of the utilization of a compass system. In the forward dynamics, system $AC_A$ are initialized in $\ket{\psi}_A\ket{0}_{C_A}$ and undergo $U \cdot U^\dag \otimes \id$. 
    One simply implements the interferometric protocol given in the main text with intervention unitaries $\tilde{V}$ and $\tilde{W}$, instead of $V$ and $W$, on $AC_A$ and $BC_B$ respectively. The diagram depicts the forward dynamics, however, under time-reversal symmetry, the temporal orientation can be actually reversed, without the experimenter knowing it. In other words, one cannot distinguish $(\dyad{\psi}\otimes \dyad{0},U\cdot U^\dag \otimes \id)$ from $(U\dyad{\psi}U^\dag\otimes \dyad{0},U^\dag\cdot U \otimes \id)$.
    }
    \label{fig:CompassInterfero}
\end{figure}

\begin{equation} \label{eqn:SMCompFP}
    (\cE_{B|A}\otimes \id_C) \star_{FP} (\rho_A \otimes \dyad{0}_{C_A}),
\end{equation}
where $\id_C$ is understood as the identity channel from $C_A$ to $C_B$. Now consider the interferometric protocol for this joint dynamics. (See Fig. \ref{fig:CompassInterfero}.)
The synchronization of this state over spacetime can be utilized as follows. When intervention unitaries $\tilde{V}_{AC_A}$ and $\tilde{W}_{BC_B}$ are given as
\begin{gather}
    \tilde{V}_{AC_A} = V_A \otimes \dyad{1}{0}_{C_A} + V^\dag_A \otimes \dyad{0}{1}_{C_A},\\
    \tilde{W}_{BC_B} = W_B \otimes \dyad{0}{1}_{C_B} + W^\dag_B \otimes \dyad{1}{0}_{C_B},
\end{gather}
the interference term $\cI$ is given as
\begin{align}
    \cI=& \Tr[(\tilde{V}_{AC_A} \otimes \tilde{W}_{BC_B}) (\cE_{B|A}\otimes \id_C) \star_{FP} (\rho_A \otimes \dyad{0}_{C_A})]\\
    =& \frac{1}{2}\Tr[(\tilde{V}_{AC_A} \otimes \tilde{W}_{BC_B})\qty{(\rho_A\otimes\dyad{0}_{C_A}\otimes\mds{1}_{BC_B}),(J[\cE_{B|A}]\otimes \swap_{C_AC_B})}]\\
    =&\frac{1}{2}\Tr[\qty{(\tilde{V}_{AC_A} \otimes \tilde{W}_{BC_B}),(\rho_A\otimes\dyad{0}_{C_A}\otimes\mds{1}_{BC_B})}(J[\cE_{B|A}]\otimes \swap_{C_AC_B})].
\end{align}
However, since $\tilde{V}_{AC_A}(\rho_A\otimes \dyad{0}_{C_A})=V_A\rho\otimes\dyad{1}{0}_{C_A}$ and $(\rho_A\otimes \dyad{0}_{C_A})\tilde{V}_{AC_A}=\rho V_A^\dag\otimes\dyad{0}{1}_{C_A}$ and $\Tr_{C_A}[(M_{C_A}\otimes\mds{1}_{C_B})\swap_{C_AC_B}]=M_{C_B}$ for any operator $M$ on $C_A$, we have
\begin{align}
    \cI=& \frac{1}{2}\Tr[(V_A\rho_A \otimes \tilde{W}_{BC_B}(\mds{1}_{B}\otimes \dyad{1}{0}_{C_B}) + \rho V^\dag \otimes \tilde{W}_{BC_B}(\mds{1}_{B}\otimes \dyad{0}{1}_{C_B}) (J[\cE_{B|A}]\otimes \mds{1}_{C_B})].
\end{align}
Similarly to $\tilde{V}_{AC_A}$, we have $\tilde{W}_{BC_B}(\mds{1}_{B}\otimes \dyad{1}{0})=W_B\otimes \dyad{0}_{C_B}$ and $\tilde{W}_{BC_B}(\mds{1}_{B}\otimes \dyad{0}{1})=W_B\otimes \dyad{1}_{C_B}$. Thus, after partial trace over $C_B$, it follows that
\begin{align}
    \cI=& \frac{1}{2} \qty(\Tr[ (V_A \otimes W_B) (\rho_A \otimes \mds{1}_B) J[\cE]_{B|A}] + \Tr[ (V_A^\dag \otimes W_B^\dag) J[\cE]_{B|A}(\rho_A \otimes \mds{1}_B)] )\\
    =&\Re[\Tr[(V_A\otimes W_B) \cE_{B|A} \star_L \rho_A]],
\end{align}
because $\cE_{B|A}\star_L\rho_A = (\rho_A \otimes \mds{1}_B) J[\cE]_{B|A}$ and $\qty(\Tr[ (V_A \otimes W_B) (\rho_A \otimes \mds{1}_B) J[\cE]_{B|A}])^* = \Tr[ (V_A^\dag \otimes W_B^\dag) J[\cE]_{B|A}(\rho_A \otimes \mds{1}_B)]$. Hence, by repeating the interferometry with varying $V_A$ and $W_B$, one can fully recover the left product $\cE_{B|A}\star_L \rho_A$ associated with the dynamics.

Now, when one has an ensemble of dynamics $(\rho^{(i)}, \cE^{(i)})$ with some probability distribution $p_i$ instead of a single dynamics $(\rho,$, the FP-product in \eqref{eqn:SMCompFP} becomes
\begin{equation}
    \sum_i p_i (\cE^{(i)}_{B|A}\otimes \id_C) \star_{FP} (\rho^{(i)}_A \otimes \dyad{0}_{C_A}).
\end{equation}
Then due to the linearity (in real numbers) of the interference term $\cI$, one can follow the same derviation and arrive at the expression
\begin{equation}
    \cI= \Re[\Tr[(V_A\otimes W_B) \sum_i p_i \cE_{B|A} \star_L \rho_A]].
\end{equation}

\section{Experimental feasibility of interferometric protocol} \label{SM:Feasible}

\subsection{Interferometric protocol as a standard scattering circuit}
\label{SM:_scattering_circuit}

The interferometric protocol underlying causally agnostic measurements can be implemented using the
well-established \emph{scattering circuit} (equivalently, Hadamard test) \cite{Miquel2002TomographySpectroscopy, Paz2003ProgrammableExpectation, PazRoncagliaSaraceno2004PhaseSpaceTomography} architecture that encodes an
interference term into the coherences of a single ancilla qubit called a probe quit.
Such ancilla-interferometric circuits are routinely used to access dynamical quantities such as
characteristic functions of work distributions and multi-time correlation functions
\cite{Mazzola2013Work,Pedernales2014NTime,Halpern2017JarzynskiOTOC},
and they were recently employed in an NMR demonstration of quantum causal inference via scattering circuits
\cite{liu2024NMR}.

\paragraph{Scattering circuit.}
A scattering circuit for probing quantum state $\rho$ on system $S$ is given as follows \cite{Du2006ScatteringCircuitNMR}: Consider a probe qubit \(R\) prepared in \(\ket{+}_R=(\ket{0}_R+\ket{1}_R)/\sqrt{2}\). The two interferometer arms are encoded in the computational basis of \(R\), and a unitary intervention \(V\) on a system register \(S\) is implemented conditionally on the arm via
\begin{equation}
U_V \;=\; \ket{0}\!\bra{0}_R \otimes \mds{1}_S \;+\; \ket{1}\!\bra{1}_R \otimes V_S .
\label{eq:SM_controlled_V}
\end{equation}
After the (possibly space- and/or time-distributed) black-box process has acted, a final single-qubit tomography on \(R\) yields the interference term. In the simplest calibration (input \(\ket{+}_R\), no additional phases), the reduced state of \(R\) takes the form
\begin{equation}
\rho_R \;=\; \frac{1}{2}\Big(\mds{1}_R + \mathrm{Re}(\cI)\,\sigma_x^{(R)} + \mathrm{Im}(\cI)\,\sigma_y^{(R)}\Big),
\label{eq:SM_rhoR}
\end{equation}
so that
\begin{equation}
\mathrm{Re}(\cI)=\langle \sigma_x^{(R)}\rangle,\qquad
\mathrm{Im}(\cI)=\langle \sigma_y^{(R)}\rangle,
\label{eq:SM_ReIm}
\end{equation}
where the expectation values are estimated from repeated experimental shots.
(Experimentally, \(\langle\sigma_{x,y}\rangle\) are obtained by standard basis rotations followed by a \(\sigma_z\)
measurement of the ancilla.)

\paragraph{Matching to QSOST readout.}
In the spatially separated case, conditional interventions \(V_A\) and \(W_B\) yield \(\cI=\mathrm{Tr}\!\left[(V_A\otimes W_B)\rho_{AB}\right]\).
In the temporally ordered (dynamical) case, implementing \(V_A\) before and \(W_B\) after a channel \(\mathcal{E}:A\to B\) yields \(\cI=\mathrm{Tr}\!\left[W_B\,\mathcal{E}(V_A\rho_A)\right]\), equivalently \(\cI=\mathrm{Tr}\!\left[(V_A\otimes W_B)(\mathcal{E}\star_L\rho_A)\right]\).
The same hardware primitive, the ancilla-controlled application of local unitaries, therefore realizes the interferometric characterization of QSOSTs with no additional assumptions about the causal relation between the spacetime regions, beyond the fact that the implemented interaction is of the interferometric (commuting-on-\(R\)) form \eqref{eq:SM_controlled_V}.
The causally agnostic condition is enforced structurally: each local interaction with the probe is a controlled unary hence diagonal in the arm basis of \(R\). Hence, all such interactions commute on \(R\), and the extracted interference term \eqref{eq:SM_ReIm} is invariant under any ordering of the spacetime-localized interactions, which is precisely the operational content of causal agnosticism.

\subsection{Platform requirements and realizations with current technology}
\label{SM:_platforms}

At the hardware level, the protocol requires only:
(i) one ancilla qubit (the arm register \(R\)) with single-qubit preparation and measurement;
(ii) the ability to implement controlled unitaries between \(R\) and the system degrees of freedom that realize
the desired interventions \(V_A, W_B\) at the relevant spacetime locations;
and (iii) coherence of \(R\) over the circuit depth (which is typically shallow, as \(R\) interacts only through
controlled gates and is measured once).

These requirements are satisfied in multiple leading quantum platforms and indeed the scattering circuit technique has been already experimentally implemented for various purposes:

\paragraph{NMR.}
NMR naturally implements scattering circuits. Controlled unitaries are synthesized using standard pulse sequences and spin-spin couplings, while the real and imaginary parts of the ancilla coherence are accessed via quadrature detection of the transverse magnetization. A recent experiment demonstrated quantum causal inference using NMR scattering circuits
\cite{liu2024NMR}, which is closely aligned with the ancilla-interferometric primitive \eqref{eq:SM_controlled_V}.
This makes NMR a direct near-term candidate for implementing causally agnostic interferometric QSOST tomography in small Hilbert spaces.

\paragraph{Trapped ions.}
Trapped-ion quantum processors provide high-fidelity single- and two-qubit gates (including native entangling gates)
and high-efficiency ancilla readout via fluorescence detection, enabling controlled-unitary compilation and
ancilla tomography at scale.
For an overview of available operations and proof-of-principle experiments in this platform, see
Ref.~\cite{BlattRoos2012}.
In this setting, the ancilla \(R\) can be a dedicated ion, and \(V_A,W_B\) can be compiled into sequences of
entangling gates and local rotations in a standard manner. See Ref.~\cite{An2015JarzynskiIon} for an interferometry-based thermodynamic test with a trapped ion system.

\paragraph{Neutral atoms / cold atoms.}
Cold-atom platforms offer long-lived internal (hyperfine) states and flexible control via Ramsey-type sequences,
with multi-qubit gates available in architectures such as optical tweezers (e.g., via strong interactions).
As a broad overview of current capabilities for coherent control and quantum simulation, see
Ref.~\cite{BlochDalibardNascimbene2012}.
Here, the ancilla \(R\) may be encoded in an internal state (Ramsey interferometry) or in a dedicated ancilla atom,
and controlled interventions can be realized using the platform’s native entangling gate set. See Refs.~\cite{Bluvstein2022NeutralAtomProcessor, Cerisola2017WorkMeterAtomChip} for examples of collision circuit realization using neutral atoms.

\paragraph{Superconducting circuits.}
Superconducting qubit devices support fast high-fidelity two-qubit gates (e.g., CZ- or iSWAP-family),
high-bandwidth ancilla readout, and straightforward compilation of controlled unitaries.
A detailed review of hardware capabilities and readout/gate performance is given in Ref.~\cite{Kjaergaard2020}.
In this platform, the ancilla \(R\) is a transmon qubit; tomography of \(R\) is routine; and controlled interventions
are compiled directly in the device-native gate set. See Ref.~\cite{OftelieCampisi2025OpenWorkQST} for an example of a scattering circuit on a superconducting platform.

\paragraph{Photonic implementations.}
In photonics, the two interferometer arms can be literal spatial modes (Mach--Zehnder-type) or internal modes (e.g., polarization/time-bin), and local unitaries \(V\) are realized by linear-optical elements acting on one arm.
Ancilla tomography corresponds to standard interferometric fringe readout and/or polarization analysis. See Refs.~\cite{Lee2013NonlinearFunctionalsPhotonics, Starek2018FredkinNPJQI} for implementation of controlled-SWAP gates and interferometry on photonic systems. For an overview of photonic quantum information processing capabilities and technology, see
Ref.~\cite{Flamini2019}.

\medskip
In all platforms above, the protocol does \emph{not} require indefinite causal order or exotic resources:
causal agnosticism is achieved by choosing the interaction form (controlled operations that commute on \(R\)),
so that the extracted interference term is invariant under the (possibly unknown) ordering of spacetime-localized
interactions.

\subsection{Experimental realization of the temporal compass system}
\label{SM:_compass}

The temporal compass construction adds a single auxiliary qubit \(C\) that co-propagates from the time of \(A\)
to the time of \(B\) through (approximately) an identity channel.
Operationally, this means that \(C\) is stored coherently between the two interventions, i.e., it functions as a
quantum memory degree of freedom.
This is experimentally natural in essentially all major platforms: one can simply preserve the state of a designated
qubit/ion/atom mode between the two intervention events without knowing their causal structure.

\paragraph{Compass-assisted interventions.}
The compass protocol uses joint unitaries between the system and \(C\) of the form
\begin{equation}
\tilde V_{A C_A} = V_A \otimes \ket{1}\!\bra{0}_{C_A} \;+\; V_A^\dagger \otimes \ket{0}\!\bra{1}_{C_A},
\qquad
\tilde W_{B C_B} = W_B \otimes \ket{0}\!\bra{1}_{C_B} \;+\; W_B^\dagger \otimes \ket{1}\!\bra{0}_{C_B},
\label{eq:SM_compass_gates}
\end{equation}
which can be implemented using standard controlled-unitary primitives plus single-qubit flips on \(C\). A convenient decomposition is obtained by noting that \(\tilde V\) maps \(\ket{\psi}_A\ket{0}_C \mapsto (V_A\ket{\psi}_A)\ket{1}_C\) and \(\ket{\psi}_A\ket{1}_C \mapsto (V_A^\dagger\ket{\psi}_A)\ket{0}_C\).
Thus \(\tilde V\) can be compiled as:
\begin{enumerate}
    \item Apply a controlled-\(V_A\) conditioned on \(\ket{0}_C\)
    \item Apply a controlled-\(V_A^\dagger\) conditioned on \(\ket{1}_C\)
    \item Apply an \(X\) gate on \(C\).
\end{enumerate}
 The same compilation applies to \(\tilde W\). In the common tomography-friendly choice where \(V_A\) and \(W_B\) are taken from a Hermitian unitary basis (e.g., Pauli operators for qubits), one has \(V_A^\dagger=V_A\) and \(W_B^\dagger=W_B\), and the compass gates simplify to the particularly simple form
\begin{equation}
\tilde V_{A C_A}= V_A \otimes X_{C_A},\qquad \tilde W_{B C_B}= W_B \otimes X_{C_B},
\label{eq:SM_compass_simplify}
\end{equation}
which is directly native/efficient in essentially all gate-based platforms. However, note that when one wants to substitute $V$ and $W$ with $iV$ and $iW$ in the interference term, one should apply
\begin{equation}
\tilde V_{A C_A}= V_A \otimes Y_{C_A},\qquad \tilde W_{B C_B}= -W_B \otimes Y_C,
\label{eq:SM_compass_simplify2}
\end{equation}
not simply $iV_A\otimes X_C$ and $iW_B \otimes X_C$ because the definition of $\tilde{V}_{AC_A}$ and $\tilde{W}_{BC_B}$ is sensitive to a phase factor.

\paragraph{Readout and overhead.}
The compass protocol depicted in Fig. \ref{fig:CompassInterfero} still a case of scattering circuit. Relative to the compass-free protocol, the only overhead is (i) one additional memory qubit \(C\) whose coherence must persist between the two intervention times, and (ii) replacing \(V_A,W_B\) by two-qubit gates \(\tilde V,\tilde W\). Hence, the compass requirement is naturally satisfied in platforms with long coherence times (NMR, trapped ions, neutral atoms) and remains feasible in fast-gate platforms (superconducting circuits), provided the time separation between \(A\) and \(B\) is within the available memory coherence window (or implemented using a dedicated longer-lived memory mode). In fact, controlled-two qubit unitary gates such as the controlled-SWAP gate have been already employed in many implementation reports \cite{liu2024NMR, Bluvstein2022NeutralAtomProcessor, Lee2013NonlinearFunctionalsPhotonics, Starek2018FredkinNPJQI}, we can conclude that the compass protocol is readily implementable with the current technology.

\section{Markovianity and quantum state over time} \label{SM:Markov}
In Refs.~\cite{pollock2018operationalmarkov, pollock2018completeframework}, Markovianity of a multi-time quantum process was given in terms of the \emph{process tensor}, a tensor that encodes the temporal correlation by outputting an output state or a measurement probability for any sequence of control operations.
Its size scales as $O(d^{4n})$ when $d$ is the dimension of each local system and $n$ is the number of time steps. However, as it turns out in Ref. \cite{pollock2018completeframework}, Markovianity of a multi-time process is equivalent to that the process is simply a concatenation of CPTP maps and an initial state prepared at the beginning of the process. In Ref. \cite{lie2024unique}, it was shown that the same condition can be expressed in terms of QSOTs. In other words, a multi-time process is Markovian if the associated QOST is factorizable, i.e., there exists a chain of CPTP maps $\cE_1,\cE_2,\dots, \cE_n$ and an initial state $\rho$ such that the QSOT of this process is given as
\begin{equation}
    \cE_n \star \qty( \cE_{n-1} \star \qty( \cdots \star \qty(\cE_1\star \rho))).
\end{equation}

Hence, one can conclude that one does not need the whole process tensor (or equivalently the process matrix) when the goal is only to decide Markovianity of the given process. This equivalence not only reduces mathematical complexity but also provides a practical method of testing Markovianity. One only has to recover the QSOT associated with the process through either interferometry or weak measurement and test the factorizability, which can be systematically done by the following protocol when the given QSOT is $\rho_{A_0A_1\dots A_n}$, under the assumption that each marginal state $\rho_{A_i}$ is full-rank:
\begin{enumerate}
    \item Calculate $\rho_{A_0}=\Tr_{A_1\dots A_n}[\rho_{A_0A_1\dots A_n}]$.
    \item Calculate $\rho_{A_0A_1}=\Tr_{A_2\dots A_n} [\rho_{A_0A_1\dots A_n}]$.
    \item Calculate either $\bar{\rho}_{A_0A_1}=\rho_{A_0}^{-1}\rho_{A_0A_1}$ (under time-reversal asymmetry) or $\bar{\rho}_{A_0A_1}=(\Theta_{\rho_{A_0}}^{-1}\otimes\id_{A_1})(\rho_{A_0A_1})$ (under time-reversal symmetry) where $\Theta_\rho(M)=\qty{\rho,M}/2$.
    \item Calculate $\cE_1=J^{-1}[\bar{\rho}_{A_0A_1}]$, where $J$ is the \jami isomorphism.
    \item Similarly calculate $\bar{\rho}_{A_{i-1}A_i}$ and $\cE_i=J^{-1}[\bar{\rho}_{A_{i-1}A_i}]$ for $i=2,\dots n$.
    \item Test if each $\cE_i$ is CPTP. If they are all CPTP, then compare $\rho_{A_0A_1\dots A_n}$ with $\cE_n \star \qty( \cE_{n-1} \star \qty( \cdots \star \qty(\cE_1\star \rho)))$. If they are the same, then the QSOT is factorizable.
\end{enumerate}
The assumption of full-rankedness of each marginal state is extremely weak, since even an infinitesimal error can make them full-rank. When some of $\rho_{A_i}$ is not full-rank, by using the technique given in Appendix D of \cite{Song2025CausalExplain}, one can still use the protocol above to test factorizability.

\section{POVMs over time}
We remark that, under time-reversal symmetry, a temporal generalization of POVMs can be constructed, based on the interferometer’s measurement probabilities \eqref{eqn:intertheta}, in analogy with Born’s rule for conventional density operators, as follows:
\begin{equation}
    \Pr(\pm)=\Tr[\, M_\pm \,  (\cE \star_{FP} \rho)\,],
\end{equation}
where $M_\pm$, defined as the Hermitian part of  $\cS_\pm \mds{1} + 2\cA_\pm (V_A\otimes W_B)$, forms a POVM $\qty{M_+, M_-}$ on $AB$ as $\cS_+ + \cS_- = 1$ and $\cA_+ + \cA_-=0$ with $|\cA_\pm|\leq 1/4$. To the best of our knowledge, this is the first case of directly implementable global POVMs over time with consistent statistical interpretation.


\end{document}